\documentclass[11pt]{article}
\usepackage{graphicx}
\usepackage{amssymb}
\usepackage{amscd}
\usepackage{mathrsfs}
\usepackage{longtable,lscape}
\usepackage{amsthm}
\usepackage{amsfonts}
\usepackage{amsmath}
\usepackage{bbm}
\usepackage{float}
\usepackage{url}
\usepackage{hyperref}
\usepackage[margin=0.5cm,font=footnotesize]{caption}
\usepackage{lineno}
\usepackage{appendix}
\usepackage{subfig}

\begin{document}
\title{\bf The Origin of Quantum Mechanical Statistics: Some Insights from Research on Human Language}
\author{Diederik Aerts$^*$, Jonito Aerts Arg\"uelles$^*$, Lester Beltran$^*$, \\ Massimiliano Sassoli de Bianchi\footnote{Center Leo Apostel for Interdisciplinary Studies, 
        Vrije Universiteit Brussel (VUB), Pleinlaan 2,
         1050 Brussels, Belgium; email addresses: diraerts@vub.be,jonitoarguelles@gmail.com,lestercc21@yahoo.com,autoricerca@gmail.com} 
        $\,$ and $\,$  Sandro Sozzo\footnote{Department of Humanities and Cultural Heritage (DIUM) and Centre CQSCS, University of Udine, Vicolo Florio 2/b, 33100 Udine, Italy; email address: sandro.sozzo@uniud.it}              }

\date{}

\maketitle
\begin{abstract}
\noindent 
Identical systems, or `entities', are `indistinguishable' in quantum mechanics (QM), and the `symmetrization postulate' rules the possible statistical distributions of a large number of identical quantum entities. However, a thorough analysis on the historical development of QM attributes the origin of quantum statistics, in particular, `Bose--Einstein statistics', to a lack of statistical independence of the micro-states of identical quantum entities. We have recently identified Bose--Einstein statistics in the combination of words in large texts, as a consequence of the `entanglement' created by the meaning carried by words when they combine in human language. Relying on this investigation, we put forward the hypothesis that entanglement, hence the lack of statistical independence, is due to a `mechanism of contextual updating’, which provides deeper reasons for the appearance of Bose--Einstein statistics in human language. However, this investigation also contributes to a better understanding of the origin of quantum mechanical statistics in physics. Finally, we provide new insights into the `intrinsically random behaviour of microscopic entities' that is generally assumed within classical statistical mechanics.
\end{abstract}

\section{Introduction\label{intro}}

The foundations of thermodynamics, as prescribed by classical statistical mechanics, rely on the hypothesis that, at a microscopic level, physical systems, or `entities', behave (i) randomly and (ii) independently from each other \cite{huang1987}. A thorough analysis of the historical development of quantum mechanics (QM) reveals that the hypothesis of statistical independence started to be questioned during early quantum theory (1900--1925) and  led to the birth of quantum mechanical statistics, in particular, Bose--Einstein statistics \cite{klein1961,howard1990,darrigol1991,monaldi2009,perezsauer2010,gorroochurn2018}. However, this was done in a way that, in our opinion, is neither straightforward nor completely understood yet.

Historians and philosophers of science have carefully reconstructed early quantum theory  through articles and personal correspondence and generally agree that Einstein's 1905 article on the photoelectric effect \cite{einstein1905} can be considered as the first genuine article on the quantum nature of physical entities, more than Planck's 1900 article on the black body radiation \cite{planck1900}. Indeed, Einstein was much more explicit about `quantization' than Planck's original intention, and he directly assumed the existence of what he called `quanta of light', which we now call `photons' \cite{klein1961}. At the time, Ehrenfest, a student of Boltzmann and close friend of Einstein, was in favour of his idea of quanta of light as entities in their own right, an idea that was not shared by most of the scientific community.

Both Wien and Planck had followed the path of Boltzmann's thermodynamics, at a time when the so-called `ultraviolet catastrophe' of the classical theory of radiation was still not known.\footnote{The expression `ultraviolet catastrophe' was introduced by Ehrenfest only in 1911 \cite{ehrenfest1911}.} However, both Einstein and Ehrenfest identified a fundamental problem in Planck's law, namely, the latter seemed to violate the request of statistical independence of the micro-states of different photons. This, though, is a feature of the Boltzmann view of statistical mechanics, shared by all protagonists, Einstein, Ehrenfest, but also Planck, which also follows as a consequence of the  Maxwell--Boltzmann distribution defining the governing statistics. The latter result was rigorously proved by Ehrenfest in 1911, namely, if one assumes statistical independence, together with Einstein's hypothesis of quanta of light, then one should find back Wien's law, hence a statistical behaviour of Maxwell--Boltzmann-type, rather than Planck's law, which incorporates a statistical behaviour of Bose--Einstein type \cite{monaldi2009,ehrenfest1911}. In particular, Einstein, who was well aware of the scientific literature of the time, was the first to realize that a `certain type of collapse occurs in the ultraviolet regime', as it is in this spectral region that photons primarily exhibit their quantum nature.

In 1924, Bose worked out an alternative derivation of Planck's radiation law \cite{bose1924}, and Einstein immediately followed Bose's method, applying it to a gas of atoms or molecules \cite{einstein1924,einstein1925a,einstein1925b}.  Einstein also observed, and this worried him, that given this lack of statistical independence, it does seem as if there is a `mysterious force that somehow makes the photons behave in a statistically dependent way'. These studies were appreciated by Schrödinger \cite{schrodinger1926a}, but also marked the departure from Ehrenfest's view of statistical mechanics. In addition, the scientific community was ready to welcome the birth of QM, and these seminal ideas of the pioneers were thus incorporated in an abstract formalism that uses the mathematics of Hilbert spaces.

In modern QM, quantum mechanical statistics, namely, Bose--Einstein and Fermi--Dirac statistics, follow from two postulates, the `tensor product postulate', which rules the combination, or composition, of quantum physical entities, and the `symmetrization postulate', which rules the possible states of identical quantum entities, bosons and fermions. According to these postulates, identical quantum entities are `completely indistinguishable from a physical point of view', and it would be this complete indistinguishability that is responsible for the lack of statistical independence identified by the pioneers. However, this formal treatment of identical quantum entities, though mathematically coherent and physically fruitful, raises long-standing epistemological and empirical issues, which are not sorted yet (see, e.g., \cite{dieks2023} and references therein).  

Coming to our research, which in this article we intend to relate to the issues we have just evoked, it 
has moved from the above epistemological and mathematical foundations of QM to the development of a quantum mechanical framework for human cognition. More specifically, we have been interested in the mathematical modelling of how people combine and exchange words and concepts through language and how new entities of meaning are formed through the combination of concepts \cite{aertssozzo2016,aertssassolisozzo2016,pisanosozzo2020,aertssassolisozzoveloz2021}. Indeed, growing empirical evidence has convincingly demonstrated in the last three decades that high-level cognitive processes involving perception, language, judgements and decisions, are not well represented in the mathematical formalisms originally conceived to deal with classical entities, e.g., Boolean algebras, Kolmogorovian probabilities, fuzzy sets, and complexity theory. On the contrary, the use of the mathematical formalism of quantum theory in Hilbert space has been successful as a modelling, predictive and explanatory framework to represent entities in cognitive-linguistics domains (see, e.g.,   \cite{aertsaerts1995,vanrijsbergen2004,aerts2009a,pothosbusemeyer2009,khrennikov2010,busemeyerbruza2012,aertsbroekaertgaborasozzo2013,aertsgaborasozzo2013,havenkhrennikov2013,kwampleskacbusemeyer2015,
dallachiaragiuntininegri2015a,dallachiaragiuntininegri2015b,melucci2015,aertssozzoveloz2016,blutnerbeimgraben2016,broekaertetal2017,aertsetal2018a} and references therein). The reason is that quantum mechanical structures, better than classical structures, are able to cope with 
features such as `intrinsic uncertainty', `contextuality', `emergence', `indeterminism', and `superposition', and there is growing evidence that these features are also present in human cognition, hence they are not peculiar of micro-physical entities, as is usually believed according to the standard view underlying QM. 

More recently, we have investigated aspects of applied cognition, e.g., computational linguistics, natural language processing and information retrieval, where the meaning content of words (thus, the corresponding concepts) and texts is recovered through digital instruments, as search engines, corpora of documents and other information retrieval systems. This investigation has allowed us to  identify Bose--Einstein statistics in the combination of words, and the corresponding concepts, in texts produced by human language. We have  proved through various empirical examples, that, if one considers a large text, which can be treated as a combination of several words, then this behaves at a statistical level as a `quantum mechanical gas of bosons'. To this end, we have attributed `energy levels' to words, in an inverse proportion to their frequency of appearance in the text. Then, we have proved that the distribution of these energy levels across words does not obey Maxwell--Boltzmann statistics but, rather, Bose--Einstein statistics. We have introduced the term `cogniton' as the fundamental quantum of cognition, in such a way that the meaning content of a text can be considered as a `quantum mechanical Bose--Einstein gas of cognitons'. Finally, we have identified a `mechanism of meaning dynamics' among the words appearing in a text, which makes more  frequent, thus lower energy, words to be more likely to appear in the text than less frequent words, 
and this behaviour closely resembles a `Bose--Einstein condensate at a temperature close to absolute zero' \cite{aertsbeltran2020,beltran2021,aertsbeltran2022a,aertsbeltran2022b}.
 
In the present article, we intend to investigate a possible connection between the lack of statistical independence of the micro-states of cognitons, which occurs because Bose--Einstein is the prevailing statistics in human language, and the notion of `meaning' present in texts produced in human language. We will also show that this lack of independence caused by the presence of Bose--Einstein statistics is deeply connected with the phenomenon of entanglement as it appears in cognitive-linguistic domains.

Recently, we have attributed the formation of entanglement, in both physics and cognitive-linguistic domains, to a `mechanism of contextual updating' \cite{ijtp2023,philtransa2023}. If one writes a text, each time one adds a new word, the new word will depend on the meaning of all the others. Equivalently, each time a word is added to a text, the word `contextually updates the meaning content of the entire text', because the new word not only fits meaning-wise the entire context of the text, but also gives, by being added, new meaning to the same. Later in the article, we will illustrate that this mechanism of contextual updating brings each word into a relationship of entanglement with the other words, and in turn also introduces the lack of independence of the separate words, that is, the fact that words tend to clump together, revealing a Bose--Einstein behaviour.  

Returning now to physics, note that Einstein was deeply convinced that Planck's law, not Wien's law, was the correct one, which is why he supported Bose's approach, although he lost at that point the support of Ehrenfest, who was instead heavily critical of Bose’s method.\footnote{Ehrenfest expressed his criticism towards Bose's method to a fellow physicist, Abram Joff\'e, in a letter dated 9 October 1924: ``Precisely now Einstein is with us. We coincide fully with him that Bose's disgusting work by no means can be understood in the sense that Planck's radiation law agrees with light atoms moving independently'' (see \cite{moskovchenkofrenkel1990}, pp. 171--172). \label{ehrenfestcritics}} Einstein's ideas made it possible for him to predict the phenomenon of Bose--Einstein condensation at very low temperatures, i.e. all elementary constituents tend to occupy the same micro-state, namely, the state of minimal energy. As we have mentioned, Einstein wondered what was the nature of the mysterious force at the origin of this behaviour and the corresponding 
violation of statistical independence of the micro-states of identical photons. Though there is currently no final answer to this question, 
we believe that the identification of a Bose--Einstein quantum statistics in human language might provide some valuable insights. 
More specifically, we consider it possible that the mysterious force that brings together identical photons can somehow be related 
to the physical counterpart of the `force of meaning' that is active in a text that tells a story. Indeed, it can be shown, and we will do so in the remainder of this article, that the statistical dependence of micro-states, which manifests itself as a consequence of the presence of Bose--Einstein statistics, is of such a nature as to be compatible with a statistical dependence that occurs as a consequence of a mechanism of contextual updating carried by meaning. 

The considerations above also suggest that there is no statistical independence at the micro-physical level, hence this hypothesis, which is at the basis of classical statistical mechanics, may not hold in the micro-world.  

Our analysis indicates that there can be a powerful structural analogy between `quantum behaviour' and `cognitive behaviour'. This was already clear in the development of `quantum cognition', where the cognitive domain was modelled through quantum mathematics. But it is also possible to look at things the other way, noting that micro-world entities become more intelligible when interpreted as carrier of meaning (obviously, a different meaning than that of human conceptual entities). This last statement certainly would need to be extensively explored and justified, but that would be beyond the scope of this article. We simply mention here that the possibility that the microscopic world possesses a nature similar to that of the human conceptual world has been explored in depth by our group, in what we have termed the `conceptuality interpretation of quantum mechanics' \cite{aerts2009b,aertsetal2020}.

Finally, we observe that the modern formal treatment of identical entities in QM needs further justification from a physics perspective. Indeed, the symmetrization postulate, which entails the complete indistinguishability of identical physical entities, in some way forces onto the system the mathematics of Hilbert space in the absence of a well-understood physical motivation for its validity. As such, the postulate should be considered, at least provisionally, as a mathematical artifice that can be used for all practical purposes rather than as a physically justified procedure. In this regard, it is worth mentioning that it is the symmetrization postulate that determines by rule the entanglement of the micro-states that is connected with the violation of statistical independence in quantum statistical mechanics. It is however not clear why the correct micro-state should be a pure entangled state and not for example  a statistical mixture. 

For the sake of completeness, we sketch the content of the present paper in the following.

In Section \ref{qm}, we summarize 
the standard mathematical and conceptual arguments that enable to represent identical entities in QM as a physical theory. We also emphasize some long-standing epistemological difficulties of this mathematical treatment.

In Section \ref{example}, we establish a connection between identical entities, indistinguishability and statistical independence, using a simple example taken from human cognition.

In Section, \ref{early}, we report the current and commonly accepted historical account of the birth of quantum mechanical statistics, in particular, Bose--Einstein statistics. We stress that Einstein considered a serious hindrance the lack of statistical independence of the micro-states and mused about the existence of some mysterious force among identical photons that could cause it. 

In Section \ref{language}, we review and sharpen the identification of Bose--Einstein statistics in human language and 
analyse a possible explanation of the lack of statistical independence in terms of contextual updating.

In Section \ref{temperature}, we establish a connection between temperature and randomness at the level of `words and their place' in a text. More specifically, randomness is introduced by distributing words at random in the places they occupied in a text that tells a story. This means that we introduce the physical equivalent of the notion of `heat' in our thermodynamics for human language and cognition. In this way, we draw the lines for a further elaboration of a thermodynamic theory of human language and cognition, whose details will be published in \cite{temperature2024}.

In Section \ref{hypothesis}, we put forward the theoretical hypothesis that the mechanism of contextual updating could also be at the basis of the quantum behaviour of identical physical entities. We also analyze the role that `meaning' plays in the structure and dynamics of texts of human language and what connection can be made with Einstein's intuition about the presence of a strange force as the cause of the lack of statistical independence of the micro-states of photons in Planck's radiation law.

\section{Identity and indistinguishability in quantum mechanics\label{qm}}
The notions of identity and indistinguishability play a fundamental role in physics and have both epistemological and empirical relevance. 

We review in this section the mathematical representation of identical physical entities in Hilbert space, as presented in modern manuals of QM (see, e.g., \cite{huang1987,eisbergresnick1985}). This formal treatment automatically incorporates the physical indistinguishability of identical entities in QM, but also raises long-standing epistemological problems, which are also briefly sketched (see, e.g., \cite{dieks2023} and references therein). We will limit ourselves to consider the situation of two identical entities, for the sake of simplicity. However, the conclusions we reach in this section also hold in the general case of a large number of identical entities.

In both classical mechanics and QM, physical entities have a set of intrinsic properties, e.g., all electrons, photons and quarks, have the same (rest) mass, spin, electric charge, etc. This is why we call `identical' two electrons or two photons. An electron and a positron are instead not identical, because they have, e.g., different charge. In this regard, there is a fundamental difference in the way two identical entities can be `distinguished' in classical mechanics and QM. 

Indeed, let us consider two identical physical entities described by classical mechanics which, at a given time, occupy two different regions of space. Then, the dynamical equations of motion guarantee that each of the two entities follows a well-defined `trajectory' which allows us to distinguish them from each other. In other words, though identical, two classical physical entities can always be distinguished. The situation is radically different in QM and one can resort to various theoretical arguments, e.g., Heisenberg's uncertainty principle which forbids the introduction of the notion of trajectory, the possible overlapping of initially separated wave functions due to the interaction between the entities, etc., and one typically maintains that there is no physical process which allows to distinguish two identical physical entities: they are `physically completely indistinguishable' in QM \cite{huang1987,eisbergresnick1985}. 

Coming to the mathematical formulation in Hilbert space, to formally deal with identical physical entities and their indistinguishability, one introduces in QM a new postulate, the `symmetrization postulate' (also known as the `exchange symmetry principle'), in addition to the `tensor product postulate', which rules the composition of quantum entities. The symmetrization postulate requires that the state vector $|\psi(1,2)\rangle$ of two identical entities, labelled $1$ and $2$, is either `symmetric' or `anti-symmetric' with respect to the permutation $1 \leftrightarrow 2$, that is, one has either $|\psi(2,1)\rangle=+|\psi(1,2)\rangle$ (symmetric case) or $|\psi(2,1)\rangle=-|\psi(1,2)\rangle$ (anti-symmetric case). The `spin-statistics theorem', derivable within quantum field theory \cite{pauli1940}, then guarantees that physical entities with integer spin, $0$, $1$, $2$, \ldots, also called `bosons', e.g., photons, are described by a symmetric state vector, while physical entities with semi-integer spin, $1/2$, $3/2$, $5/2$, \ldots, also called `fermions', e.g., electrons, are described by an anti-symmetric state vector. 

To make the above more concrete, let $1$ and $2$ be two identical physical entities, associated with the Hilbert spaces $\mathscr{H}(1)$ and $\mathscr{H}(2)$, respectively. Moreover, let the unit vectors $|\psi_{\alpha}(1)\rangle$ and $|\psi_{\beta}(1)\rangle$ represent two possible states of entity $1$ and suppose that $|\psi_{\alpha}(1)\rangle$ and $|\psi_{\beta}(1)\rangle$ are orthogonal. Analogously, let the unit vectors $|\psi_{\alpha}(2)\rangle$ and $|\psi_{\beta}(2)\rangle$ represent two possible states of entity 2 and again suppose that $|\psi_{\alpha}(2)\rangle$ and $|\psi_{\beta}(2)\rangle$ are orthogonal. According to the tensor product postulate, the unit vectors $|\psi_{\alpha}(1)\rangle |\psi_{\alpha}(2)\rangle$, $|\psi_{\alpha}(1)\rangle |\psi_{\beta}(2)\rangle$, $|\psi_{\beta}(1)\rangle |\psi_{\alpha}(2)\rangle$ and $|\psi_{\beta}(1)\rangle |\psi_{\beta}(2)\rangle$ should represent possible states 
of the composite entity made up of entity $1$ and entity $2$ (we omit the tensor product symbol $\otimes$ in the case of vectors, for the sake of simplicity) in the tensor product Hilbert space $\mathscr{H}(1) \otimes \mathscr{H}(2)$. However, the symmetrization postulate forbids that these unit vectors represent genuine physical states, due to indistinguishability, and instead requires that the state vector of a composite entity made up of two fermions is the entangled state represented by the anti-symmetric unit vector
\begin{equation} \label{fermionstate}
|\psi_{A}(1,2)\rangle= \frac{1}{\sqrt{2}} \Big [ |\psi_{\alpha}(1)\rangle |\psi_{\beta}(2)\rangle-|\psi_{\beta}(1)\rangle |\psi_{\alpha}(2)\rangle \Big ]
\end{equation}
whereas the state vector of a composite entity made up of two bosons is the entangled state represented by the symmetric unit vector
\begin{equation} \label{bosonstate}
|\psi_{S}(1,2)\rangle= \frac{1}{\sqrt{2}} \Big [ |\psi_{\alpha}(1)\rangle |\psi_{\beta}(2)\rangle+|\psi_{\beta}(1)\rangle |\psi_{\alpha}(2)\rangle \Big ] 
\end{equation}
In both the fermion and the boson case, an immediate physical consequence is obtained if we set $\alpha=\beta$. Indeed, Equation (\ref{fermionstate}) then entails that $|\psi_{A}(1,2)\rangle=0$, that is, two fermions cannot occupy the same single-entity state, or `micro-state', a result known as the `Pauli exlusion principle'. On the contrary, if we set $\alpha=\beta$ in Equation (\ref{bosonstate}), we get $|\psi_{S}(1,2)\rangle=\sqrt{2} |\psi_{\alpha}(1)\rangle |\psi_{\alpha}(2)\rangle$, whence $|||\psi_{S}(1,2)\rangle||^{2}=2|||\psi_{\alpha}(1)\rangle |\psi_{\alpha}(2)\rangle||^{2}$. Since $|| |\psi_{S}(1,2)\rangle||^{2}$ is connected with the probability of finding two bosons in the same micro-state labelled $\alpha$, this probability is multiplied by a factor 2 with respect to the case in which the two bosons are distinguishable, which would be connected with $|||\psi_{\alpha}(1)\rangle |\psi_{\alpha}(2)\rangle||^{2}$. This expresses the typical tendency of bosons to occupy the same micro-state (see Section \ref{example}).

The requirement of indistinguishability of identical entities in QM has dramatic consequences on their statistical properties. Indeed, let us consider the composite entity made up of a large number of identical physical entities, and suppose that the situation of an ideal gas of non-interacting entities in thermal equilibrium is satisfied, as typically assumed in statistical mechanics considerations. Then, the condition of indistinguishability implies that the average number $N(E_i)$ of entities with energy $E_i$ has the mathematical form
\begin{equation} \label{fermidiracdistribution}
N(E_i)=\frac{1}{Ae^{\frac{E_i}{B}}+1}
\end{equation}  
in the case of fermions, which is known as the `Fermi--Dirac distribution', and the mathematical form
\begin{equation} \label{boseeinsteindistribution}
N(E_i)=\frac{1}{Ae^{\frac{E_i}{B}}-1}
\end{equation}  
in the case of bosons, which is known as the `Bose--Einstein distribution', where $A$ and $B$ are physical constants. These distributions deeply differ from the `Maxwell--Boltzmann distribution' 
 \begin{equation} \label{maxwellboltzmanndistribution}
N(E_i)=\frac{1}{Ce^{\frac{E_i}{D}}}
\end{equation}  
with $C$ and $D$ physical constants, governing the statistical behaviour of a large number of distinguishable physical entities in classical statistical mechanics \cite{huang1987,eisbergresnick1985}. 

Quantum indistinguishability and the corresponding quantum mechanical statistics play a fundamental role in several macroscopic quantum effects which have been widely confirmed experimentally, such as chemical bonds, the properties of semi-conductors, Bose--Einstein condensation, superfluidity and superconductivity \cite{eisbergresnick1985,cornellwieman2002,ketterle2002,annett2005}.

However, the symmetrization postulate has also relevant epistemological implications. Indeed, let us limit ourselves to consider position coordinates and neglect for a moment spin  variables.  Let $|\psi_A(1)\rangle$ and $|\psi_B(2)\rangle$ be the unit vectors representing the states of the physical entities $1$ and $2$, 
and corresponding to two wave functions which are non-zero only in the widely separated regions $A$ and $B$, respectively. Then, 
combining Equations (\ref{fermionstate}) and (\ref{bosonstate}), the symmetrization postulate requires that the state vector of the composite entity made up of entities $1$ and $2$ has to be $\frac{1}{\sqrt{2}}(|\psi_A(1)\rangle|\psi_B(2)\rangle \pm |\psi_B(1)\rangle|\psi_A(2)\rangle)$. In both cases, we obtain a bi-orthogonal decomposition of a unit vector representing an entangled state, and it is well known from the standard interpretation of the coefficients of a bi-orthogonal decomposition, that in neither case one can say that one  entity is localized in the space region $A$ and the other entity is localized in the space region $B$, as it would be natural in the case of well separated wave functions. According to some authors \cite{dieks2023,diekslubberdink2020}, these epistemological difficulties  
make problematical the possibility of experimentally preparing a composite physical entity, made up of two identical quantum entities, in a pure state in which the component entities are spatially separated but entangled in spin, such as `EPR-Bell-type states' (see, e.g., \cite{aertsbeltran2022a,ijtp2023,philtransa2023}). Various approaches have been put forward to solve these difficulties, but there is no general consensus in the scientific community about their resolution \cite{dieks2023,diekslubberdink2020,frenchredhead1988,saunders2003,mullerseevinck2009,krause2010}.

The above considerations allow one to conclude that the symmetrization postulate provides a mathematical tool which works well to derive predictions in agreement with empirical data. However, on the one hand, it has little physical justification and, on the other hand, it raises various long-standing and not completely solved issues at an epistemological level. In addition, the condition of complete physical indistinguishability, implied by the request of symmetrization, is at odds with the way in which experimental quantum physicists deal with identical quantum entities in their laboratories, as we will see in Section \ref{hypothesis}.

\section{Indistinguishability versus statistical independence \label{example}}
We intend in this section to establish a connection between distinguishability of identical entities and statistical independence of their relevant states. We will first consider the typical situation that occurs in physics \cite{gorroochurn2018}. Then, we will see that a completely analogous connection can be made in the case of human cognition, more specifically, language \cite{aertsbeltran2022a,aertsbeltran2022b}.

Traditional derivations of both classical, Maxwell--Boltzmann, and quantum, Bose--Einstein and Fermi--Dirac, distributions employ a method of distributing particles of the same type, which play the role of identical physical entities, across baskets, which play the role of micro-states (see, e.g., \cite{huang1987,darrigol1991}). Let us consider a simple situation, namely, the distribution of two physical entities, labelled $1$ and $2$, in two micro-states, labelled  $p_1$ and $p_2$ \cite{gorroochurn2018}. We will see, by comparing the classical case with the quantum ones, that the requirement of physical indistinguishability has a dramatic impact on the independent behaviour of physical entities that is traditionally assumed at a statistical level. 

We start by the case of two entities $1$ and $2$ 
described by classical mechanics. In this case, the entities are distinguishable and one has four possible realizations, that is, (i) $1$ in $p_1$ and $2$ in $p_1$, (ii) $1$ in $p_1$ and $2$ in $p_2$, (iii) $1$ in $p_2$ and $2$ in $p_1$, and (iv) $1$ in $p_2$ and $2$ in $p_2$, and each realization is associated with a probability equal to $1/4$. In other words, the probability that both entities occupy a given micro-state is $1/4$, which is the product of the probabilities, $1/2$ and $1/2$, that each entity occupies that micro-state. In this case, we say that the entities are `statistically independent'. If one then assumes an `epistemic-only indistinguishability', as typically done in statistical mechanics, one gets three realizations, that is, (i) two entities in $p_1$, (ii) one entity in $p_1$ and the other in $p_2$, and (iii) two entities in $p_2$, associated with probabilities $1/4$, $1/2$ and $1/4$, respectively. This is at the basis of Maxwell--Boltzmann statistics, whose distribution has the form written in Equation (\ref{maxwellboltzmanndistribution}).

The situation is deeply different if the two physical entities are described by QM, hence they should be regarded as physically indistinguishable (see Section \ref{qm}). Again, suppose that each entity has a probability $1/2$ of occupying the micro-state $p_1$ and a probability $1/2$ of occupying the micro-state $p_2$.  

Let us consider the case of identical bosons, e.g., two photons. If the entities are indistinguishable, one has three possible realizations, that is, (i) two entities in $p_1$, (ii) one entity in $p_1$ and the other in $p_2$, and (iii) two entities in $p_2$, and and each realization is associated with a probability equal to $1/3$. The probability that both entities occupy a given micro-state is now $1/3$, rather than $1/4$, and the entities are no longer statistically independent. This is at the basis of Bose--Einstein statistics, whose distribution has the form written in Equation (\ref{boseeinsteindistribution}). An interesting physical effect also arises in the case of Bose--Einstein statistics, as anticipated in Section \ref{qm}, namely, the probability of two entities occupying the same micro-state is $2/3$, whereas the probability of two entities occupying different micro-states is $1/3$: bosons tend to `cluster together more often than statistical independence would predict'.   

Finally, let us investigate the case of identical fermions, e.g., two electrons. In this case, the entities are still indistinguishable but no micro-state can be occupied by two entities at the same time, hence one remains with a single realization, that is, (i) one entity in $p_1$ and the other in $p_2$, associated with a probability equal to $1$.  This is at the basis of Fermi--Dirac statistics, whose distribution has the form written in Equation \ref{fermidiracdistribution}. Also in this case, a lack of statistical independence occurs, which results in fermions tending to be more separate than independence would predict, as a consequence of the Pauli exclusion principle.

For the purposes of the present article, we will not dwell on the appearance of Fermi--Dirac statistics in human language (for the interested reader, we refer to \cite{aertsbeltran2022b}). We instead intend to clarify the differences between Maxwell--Boltzmann and Bose--Einstein statistics, in particular, with respect to the issue of statistical independence, using a simple but paradigmatic example, taken from human cognition, more specifically, language, as follows \cite{aertsbeltran2022a}.

Suppose we visit a farm which contains an equal number of cats and dogs, and we ask the farmer to choose two animals for us, each animal being a cat or a dog. Moreover, let us suppose that the farmer chooses each animal at random, that is, a cat with probability $1/2$ and a dog with probability $1/2$. Thus, the two animals and the cat/dog states are the counterpart of the two physical entities $1$ and $2$  and micro-states $p_1$ and $p_2$ above, respectively, in this example. Let us also assume that the choice of the first animal does not affect the choice of the second. Equivalently, the farmer could choose at random based on the toss of a fair coin.

It is easy to check that there is a probability $1/4$ that we are offered two cats, a probability $1/4$ that we are offered two dogs, and a probability $1/2$ that we are offered a cat and a dog. This is the counterpart of a typical Maxwell--Boltzmann situation where individual animals can be distinguished and at most the epistemic indistinguishability above is assumed. More precisely, the Maxwell--Boltzmann description is obtained in this example if we consider the farmer's choice of `a cat and a dog' as different from the choice of `a dog and a cat'. For example, if the cats and dogs are put in two different baskets, the choice `a cat in the first basket and a dog in the second basket' can be easily distinguished from the choice `a dog in the first basket and a cat in the second basket'. The probabilities associated with these two different choices are then $1/4$ and $1/4$, respectively, each probability being the product of the independent probabilities $1/2$ and $1/2$ for choosing cat or dog. This is the counterpart of the above statistical independence of the micro-states in this example. 

To recap our example, even if one does not distinguish between `cat and dog' and `dog and cat', the ensuing statistics has still the form of the Maxwell--Boltzmann distribution in Equation (\ref{maxwellboltzmanndistribution}). On the contrary, the counterpart of Bose--Einstein statistics assigns the same probability $1/3$ to `two cats', `two dogs', and `a cat and a dog'. Suppose for a moment that this Bose--Einstein  statistics emerges with a procedure of choosing the cats and dogs and placing them in a kennel. 
And let us suppose that the first animal, a cat or a dog, has already been chosen and placed in a kennel, and it is now the turn to choose the second animal. The first was chosen with probability $1/2$ being a cat and probability $1/2$ being a dog. And suppose the first animal is a cat. Since we must arrive at probability $1/3$ that two cats are chosen, choosing a second cat must be associated with a probability $2/3$, and consequently the probability of choosing a dog, after choosing a cat, is reduced to $1/3$. 
That means that the second choice takes place with drastically changed probabilities, namely $2/3$ for cat and $1/3$ for dog, than would have been the case had the first choice not been made (since the probabilities would then have been $1/2$ and $1/2$, respectively). This indeed evokes the image of a mysterious force that suddenly after the first choice with outcome cat makes it more likely to choose a cat again at the second choice (probability $2/3$) and less likely to choose a dog (probability $1/3$). 

The choice procedure we have just described, with cats and dogs being chosen one at a time, sequentially, and placed in a kennel, is however not the only one, and we now want to offer an example of a different way of choosing. Let us suppose that the animals' choice is not made by the farmer but, rather, by a child who has been promised he/she can have two pets and choose for himself/herself whether either pet is a cat or a dog. In the realm of the child's conceptual world, there is no reason a priori that `a cat and a dog' should be preferred to `two cats' or `two dogs', because in that conceptual world, either there are two kitties playing together, or two puppies playing together, or a kitty and a puppy playing together, which could easily lead to the same probabilities $1/3$, $1/3$, and $1/3$, for these three situations,   hence to Bose--Einstein statistics. Of course, there are many factors influencing the child's choice: it might be that `a cat and a dog' has a probability less than the $1/3$ value assigned by the Bose--Einstein statistics, e.g., the child may have been told that a cat and a dog could lead to problems when they grow up. But, if one considers many children, it is reasonable to expect that a deviation from Maxwell--Boltzmann statistics will occur in favour of Bose--Einstein statistics.

The considerations above provide a first important insight that, in both physics and cognitive-linguistics domains, the assumption of distinguishability (respectively, indistinguishability) is deeply connected with the assumption of statistical independence (respectively, dependence) for the relevant states of the entities considered in these domains and influences the type of statistical distribution that can be realized when several entities of the same type are considered. This will be reinforced in Section \ref{language}, where we will show that identical words in large texts generally do not distribute according to Maxwell--Boltzmann but, rather, Bose--Einstein statistics. In the next section, we instead intend to present arguments to support the idea that Bose--Einstein statistics was historically discovered exactly from the recognition that a lack of statistical independence occurs in the micro-states of identical physical entities.

\section{Violation of statistical independence in early quantum theory\label{early}}
Bose--Einstein statistics was discovered before the advent of QM, in a period historiographically known as `early quantum theory' and approximately located between 1900 and 1925. Historians and philosophers of science have carefully reconstructed the intense debate which led to the birth of the new statistics, using both scientific articles and personal correspondence as sources \cite{klein1961,howard1990,darrigol1991,monaldi2009,perezsauer2010,gorroochurn2018}. In this regard, it is interesting to sketch the different positions of the major players involved in the scientific and epistemological discussions about Bose--Einstein statistics.

The starting point of this story concerned the investigation of how a material entity that is heated begins to emit electromagnetic radiation. An entity that is particularly suited for studies of this type is the `black body', namely, an ideal, but physically realizable with very good approximation, body that absorbs all the electromagnetic radiation incident on it \cite{eisbergresnick1985}. In the last years of 1800, it was well known the empirical relationship connecting the energy density and the frequency of the electromagnetic radiation emitted by a black body heated at a given temperature. 

Modern manuals of QM typically report that the formal treatment of the interaction between matter and electromagnetic radiation within classical physics leads to a theoretical law, the `Raileigh--Jeans law' which diverges from empirical data at high frequencies, the so-called `ultraviolet catastrophe' \cite{ehrenfest1911}. It was Max Planck who proposed in 1900 what is currently accepted as the correct theoretical law of the black body radiation, in this way determining the birth of early quantum theory \cite{planck1900}. However, the idea of an ultraviolet catastrophe came only after the seminal article of Planck, who was instead inspired by the law of the black body radiation formulated by Wilhelm Wien, which showed good agreement with the empirical data available at the time \cite{wien1897}. Nonetheless, experiments were performed in 1900 which did not agree with Wien's law in the case of radiation with low frequency \cite{rubenskurlbaum1900}.

Looking at a thermodynamic description of the black body radiation, both Wien and Planck were deeply influenced by Boltzmann's statistical view of thermodynamics \cite{boltzmann1884}. Indeed, Planck derived his law of the black body radiation while trying to provide a thorough derivation of Wien's law from Boltzmann's thermodynamics. The introduction of the constant $h$, now named after him, together with the constant $k$, which Planck named after Boltzmann, was primarily intended as a rescue operation for Planck's thermodynamic theory of the electromagnetic radiation. In an almost miraculous way, Planck's law, which approximates Wien's law at high frequencies such that experiments at high frequencies could not show deviations of Wien’s law, also worked very well at low frequencies, thus showing very good agreement with the new empirical data. In modern terms, if we recall the mathematical formulation of Maxwell--Boltzmann and Bose--Einstein statistics introduced in Section \ref{qm}, we can say that Wien's law is the Maxwell--Boltzmann approximation of Planck's law, which instead holds within Bose--Einstein statistics.

The systematic and complete historical account in \cite{klein1961,howard1990,darrigol1991,monaldi2009,perezsauer2010,gorroochurn2018} of the scientific debate within early quantum  theory reveals that Planck's idea of `quantization', initially conceived as a mathematical artifice, was still only partial and quite different from the way we understand it nowadays. As a matter of fact, it was Albert Einstein, who applied Planck's radiation law to the photoelectric effect \cite{einstein1905}, to give a physical meaning to the formula $h\nu$ introduced by Planck and interpret it as the energy of a `quantum of light', later called `photon' \cite{klein1961}. 

Both Planck and Einstein solved their respective problems by looking at the way in which `particles can be distributed in baskets' (see Section \ref{example}) \cite{darrigol1991}. But, Einstein, and later Paul Ehrenfest, recognised a fundamental problem in the procedure followed by Planck. Indeed, Planck used the method of distributing identical particles into different baskets to find the distribution of energies across the elementary constituents of the electromagnetic radiation. In this way, he found a formula for the black body radiation which coincides with the formula one finds starting directly from the Bose--Einstein distribution. When Einstein studied the photoelectric effect, he attributed a particle nature to those elementary constituents of the radiation, as atoms and molecules. However, it immediately became clear that, if the idea of photons is taken seriously, then the `probabilities for the relevant micro-states did not appear to be statistically independent'. As a matter of fact, Ehrenfest later proved rigorously that, assuming statistical independence in a gas of photons, would lead to Wien's, not Planck's, law \cite{ehrenfest1911,monaldi2009}. It is also why Planck, who still relied on Boltzmann's approach, was not at all inclined to assign a physical reality to these quanta and continued to regard them as abstract elements of a specific thermodynamic view of the situation. 

Hence, the positions of Planck and Einstein (together with Ehrenfest) with respect to the interpretation of the radiation law was dramatically divergent. Planck, and with him most other physicists working on related problems, was not troubled by the lack of statistical independence, and thought this was only problematic because Einstein insisted on interpreting these quanta as entities existing in time and space. Einstein and Ehrenfest, on the contrary, considered a serious hindrance the identification of a lack of statistical independence at the level of micro-states, which they could not justify, wanting to interpret these micro-states as physical states of quanta of light, entities existing in time and space, and both looked for an alternative theoretical proposal. This is why, when Einstein learned about Bose's alternative derivation of Planck's radiation law, he enthusiastically translated Bose's article into German and submitted it to Zeischrift für Physik, where the article was published in 1924 \cite{bose1924}. Furthermore, Einstein immediately applied Bose's approach to an ideal gas of atoms or molecules in a series of articles between 1924 and 1925, which led to the identification of the phenomenon of `Bose--Einstein condensation' \cite{einstein1924,einstein1925a,einstein1925b}. 

This shift of perspective by Einstein also marked the departure from agreement with Ehrenfest, who was not at all enthusiast about Bose’s type of calculation, even being amazed that Einstein embraced it so strongly (see footnote \ref{ehrenfestcritics}) \cite{perezsauer2010,moskovchenkofrenkel1990}. In the meanwhile, Louis de Broglie had formulated his hypothesis about the wave nature of material entities \cite{debroglie1924}, which Einstein read with great attention. Indeed, in particular in \cite{einstein1925a}, Einstein saw the statistical dependence of the photons in a gas as caused by a `mutual influence’, whose nature was totally unknown but, according to Einstein, de Broglie wave-particle duality could have played a role in it \cite{monaldi2009,einstein1925a}.

Now, we have sketched in Section \ref{qm} the issues arising from the modern way of framing the description of quantum entities, bosons and fermions, as if this description followed from their complete physical indistinguishability. We stress that these issues started to appear already in the discussions among the founding fathers of QM. Indeed, the experimental realization of Bose--Einstein condensates in which these form spontaneously when the disturbing factor of heat photons is removed by sophisticated techniques of cooling \cite{cornellwieman2002,ketterle2002}, shows that the undisturbed bare state of the gases used is coherent and not optimally random as would follow from both Boltzmann's and Gibbs' thermodynamics \cite{boltzmann1878,gibbs1902}. Einstein's articles on gases, with Bose's calculation, did not go unnoticed in the scientific circles where QM was being pondered but did not get any support. We have mentioned above how negative Ehrenfest's reaction was to Einstein embracing Bose's calculation. Einstein, probably looking for support, wrote the following in a letter to Schrödinger, and it is also in this letter he calls Bose's calculation a `statistics' for the first time: ``In Bose's statistics, which I have used, the quanta or molecules are not treated as independent from one another. [\ldots] According to this procedure, the molecules do not seem to be localized independently from each other, but they have a preference to be in the same cell with other molecules. [\ldots] According to Bose, the molecules crowd together relatively more often than under the hypothesis of statistical independence'' (Einstein to Schrödinger, Berlin, 28 Feb. 1925 \cite{monaldi2009}). 

The new statistics of Bose and Einstein also came to the attention of Planck, who noted that, if the correlations between molecules were confirmed experimentally, this would entail ``a fundamental modification of the ordinary conception of the nature and mode of interaction of the molecules'' \cite{monaldi2009,planck1925}. Schrödinger also responded to the new statistics with an article sharply expressing his disapproval, in the sense that Bose's procedure represented ``a radical departure from the Boltzmann--Gibbs kind of statistics.'' Reacting to the explanation he received in Einstein's letter, he objected that he did not see ``for the time being any possibility of understanding the remarkable kind of interaction among the molecules'' by which it could be justified \cite{monaldi2009,schrodinger1925}. It is also in this article by Schrödinger that the symmetrization postulate (or exchange symmetry principle, see Section \ref{qm}) was brought forward for the first time with respect to statistics applied to quantum entities. In 1926, and even before he published the article introducing matter waves, Schrödinger wrote a second article directly on Einstein's gases. Also in this article, Schrödinger remains dismissive of the new Bose--Einstein statistics, which according to him can never be the `natural statistics' of a thermodynamic equilibrium, as it must always be the Boltzmann- or Gibbs-type of statistics with independence of the different micro-states. He also hints in this article that the wave character as put forward in de Broglie's work might bring solace relative to the situation \cite{monaldi2009,schrodinger1926a}. 

Einstein's stance had not escaped Paul Dirac either. In an unpublished manuscript, he wrote that the molecules were ``not distributed independently from one another'', so that there must be ``some kind of interaction between them'' \cite{monaldi2009,dirac1925}. In later formulations of the two quantum statistics, Bose--Einstein and Fermi--Dirac, and likewise in the standard introduction that can now be read in most textbooks, the concern about the lack of independence of the micro-states, and the question of what could be an explanation for it, disappeared in its totality. One of the originally highly critical protagonists, Erwin Schrödinger, much later, while working in Ireland having left Austria as a consequence of Nazism, gave a course on the subject of `statistical thermodynamics'. A first publication of his course took place in 1944 and a more elaborate book was written in 1952 \cite{schrodinger1989}, but Schrödinger's approach, inspired by the work of Gibbs, is not different from what is now standard in modern textbooks. In these texts, the disappearance of the lack of statistical independence is so profound that theoretical physicists are usually unaware that concerns about it, with the consequent question of what could explain it, played such an important role in the first decades of QM's development and that almost all of its protagonists reflected explicitly on it. 

So, how and when did the turnaround come to the current textbook introduction of quantum statistics, where both bosons and fermions, i.e. `all' quantum entities, are considered to be completely physically indistinguishable? Daniela Monaldi has done historical research with exactly a focus on this question. She mentions as a first root of indistinguishability the awareness of correlations between micro-states when they are stripped of heat-generated randomness, as we have explained above, with Planck's radiation law as the focus 
of research and Einstein, Ehrenfest and Planck its main protagonists.  As a second root however, she identifies the modification of the classical calculation of the entropy of the monatomic ideal gas, namely, the subtraction of a term depending on the number of possible permutations of identical particles, which is still today frequently justified as a correction required by the symmetry of identical multi-particle entities under their exchange. She also notes that, if in hindsight one can identify these two issues as two roots of indistinguishability, for the purpose of reconstructing the history of quantum statistics these roots are not the plant \cite{monaldi2009}. 
Indeed, despite superficial similarities with the modern formulation, the early studies of Planck's law did not point to the existence of an alternative type of statistics. Furthermore, no one drew any connection between the non-independence of light quanta and an exchange symmetry. This is noteworthy, since exchange symmetry was at that time becoming an issue in ideal gas theory, an issue that is considered in hindsight the second root of indistinguishability. 

Monaldi analyzes in detail passages in various articles by scientists who worked in the wake of Einstein, Ehrenfest and Planck on the hypothesis of the existence of light quanta, and where the introduction of a different than the Maxwell--Boltzmann statistics might have been at issue. She concludes that this was not the case. She also notes that the important hypothesis of exchange symmetry has never been discussed in the research of scientists who have worked on light quanta. That the notion of indistinguishability occupies such a prominent place in the standard way of presenting the two quantum statistics is thus primarily connected with the second root, the problem of entropy that does not sum as expected in the situation where gases are mixed. Especially Gibbs in his version of statistical thermodynamics paid much attention to this problem, devoting the entire second chapter of his book to it \cite{monaldi2009,gibbs1902}. Gibbs is also the one who put forward exchange symmetries as important, but his work was not directly related to the emergence of QM taking off with Planck's work. Nevertheless, because both Planck and Einstein wished to describe gases and light in an integrated way, there was regular influence from prominent figures in gas theory on the group engaged in the development of QM. Even today there is still an overlap of the group of researchers working on what has since been called the `Gibbs paradox' and the typical problems posed by QM (see, e.g., \cite{dieks2011}).

We will not elaborate further on this second root in this article since we plan to systematically investigate it in future work, examining to what extent our results related to the Bose--Einstein statistics of human languages are able to shed light on the Gibbs'  paradox. In summary, the technical form of Bose--Einstein and Fermi--Dirac statistics arose as a result of Einstein's enthusiasm for Bose's unorthodox method of calculation, but it remained unexplained why this calculation made sense. In the final section of this paper, we will complete
our critical look at the modern standard introduction with insights related to our investigation of human language. 

Coming back to Einstein, we mentioned that he embraced Louis de Broglie's matter waves, but with mixed feelings, as they allowed a possible explanation for the lack of statistical independence contained in Planck's law. Also, the de Broglie’s waves, since they are linked to individual entities, do not stand in the way of the philosophical realism that Einstein prioritized as the criterion for acceptability of an interpretation of the theory. And perhaps, though we suspect this was not at all cleared up for Einstein how that could be possible, they could explain that mysterious lack of statistical independence of photon states. We sense here, in our view, the deep intuition that one also finds in Einstein's other works, and the courage and stubbornness to stick to it, albeit in opposition to his greatest ally in the struggle for acceptance of the existence of light quanta, Ehrenfest. But Schrödinger's introduction of waves in the configuration space of the physical entities \cite{schrodinger1926b},  was for Einstein a bridge too far. Indeed, such waves cannot belong to a reality consistent with Einstein's philosophical realism, and must trigger deep problems of interpretation in connection with measurement. Einstein was also aware that such waves in configuration space would create a fundamental difficulty for the search of a possible quantum version of his general theory of relativity \cite{howard1990}.

We have seen in Section \ref{qm} that modern QM connects Bose--Einstein statistics to the indistinguishability of identical bosons, which is formally represented by the symmetrization postulate. The latter postulate forces an exchange symmetry in the states of the Hilbert space, producing through a mathematical artifice the entangled states that would be responsible of the lack of statistical independence. However, we have also stressed the epistemological difficulties arising from the symmetrization postulate. Moreover, this formal treatment leads to additional difficulties. Indeed, according to the standard interpretation of identical entities in QM, all photons would be `completely indistinguishable', even though they have different energies, e.g., a blue photon would be indistinguishable from a red photon. This assumption does not completely fit with the empirical research on how indistinguishability of photons is used in order to prepare entangled photons and treat them as qubits  in quantum computational tasks. What arises in these experiments is that it is sufficient for the photons to be `contextually indistinguishable' when they are measured for them to behave as indistinsguishable bosons (see \cite{aertsbeltran2020} and references therein). In other words, photons of different energies are generally considered by quantum experimentalists as distinguishable entities.

The epistemological and physical difficulties of standard QM in the representation of identical entities suggest that alternative theoretical proposals should be considered to explain the emergence of quantum statistics. In that regard, we have seen in the simple example presented in Section \ref{example} how the difference between Maxwell--Boltzmann and Bose--Einstein statistics in human language can be naturally explained in terms of the difference between statistical independence and statistical dependence. In addition, the identification of Bose--Einstein statistics in human language that we have anticipated in Section \ref{intro} relies on the fact that the words of a text produced by human language can be considered as different energy states of an overall entity, called the `cogniton', 
and are not considered as indistinguishable in human language. In this sense, the different words of a text play in human language the same role that the photons with different energy play in experimental QM.

It is then worth to sketch the fundamental aspects of our research on the identification of Bose--Einstein statistics in large texts produced by human language in order to check whether our approach also provides a more compelling explanation of how Bose--Einstein statistics arises in QM as a physical theory. This will be the aim of Section \ref{language}.

\section{The appearance of Bose--Einstein statistics in language\label{language}}
In this section, we summarize the theoretical scheme we have recently developed to identify Bose--Einstein statistics in human language, specifically, in the distribution of words, and their respective concepts, in large texts.

We do not expound the theoretical scheme in detail here and limit ourselves to summarize the main results, for the sake of brevity (for the full development of these ideas, see, e.g., \cite{aertsbeltran2020,beltran2021,aertsbeltran2022a,aertsbeltran2022b} and references therein). We instead aim, in this and the following section, to explain why Bose--Einstein statistics arises in human language and to reflect about the specific lack of statistical independence that Einstein identified in the behaviour of a gas of photons and how in cognitive-linguistics domains there could be a connection with a `mechanism in the dynamics of meaning' that occurs whenever words, and their respective concepts, are combined to produce a text of human language.

We preliminarily stress that we consider words as labels for concepts in our theoretical scheme, that is, they are the concepts that give the words they refer to their meaning. More precisely, we consider any concept as an abstract entity of meaning that is at a given time in a given state, and the latter is what captures the meaning content of the concept. Individual concepts can then be composed, or combined, to form more complex entities of meaning, e.g., conceptual combinations. Proceeding in this way, a written text is an entity of meaning obtained by combining the concepts that correspond to all the words that appear in the text. We have demonstrated in several articles that the meaning of a written text produced by human language relates to the meaning of the words that appear in the text in a complex way which deviates from the prescriptions of classical logical semantics (see, e.g., \cite{aertsbeltran2022a,aertsbeltran2022b,philtransa2023}).

We have seen in Section \ref{qm} that the Maxwell--Boltzmann distribution in Equation (\ref{maxwellboltzmanndistribution}) models the statistical behaviour of identical physical entities as a consequence of their distinguishability, randomness and independence. On the other side, indistinguishability makes the Bose--Einstein distribution in Equation (\ref{boseeinsteindistribution}) the correct distribution for the statistical behaviour of identical bosons. We have also seen in Section \ref{example} that indistinguishability determines, in particular, in the Bose--Einstein case, a lack of statistical independence, which Einstein finally attributed to a mysterious force that makes identical bosons to clump together in the same micro-state. However, our remarks in Section \ref{example} allow us to conclude that indistinguishability is responsible of a lack of statistical independence also in conceptual entities within human cognition.

Why shall we expect Bose--Einstein statistics to hold in conceptual combinations?  To understand this point, let us firstly consider the simple concept combination {\it Eleven Animals}. It is clear that, at a conceptual level, each one of the eleven animals is completely `identical with' and `indistinguishable from' each other of the eleven animals. On the other side, it is also clear that, in the case of `eleven physical animals', there are always differences between each of them because, as `objects' present in the physical world, they have an individuality and, as individuals with spatially localized physical bodies, they can be distinguished from each other. In the latter case, we expect Maxwell--Boltzmann statistics to apply. In the former case, we instead expect a non-classical statistics to apply as a consequence of the fact that, due to their conceptual nature, the eleven animals are intrinsically indistinguishable. We have proved in \cite{aertssozzoveloz2015} that, indeed, Bose--Einstein statistics better models the concept combination {\it Eleven Animals} as compared to Maxwell--Boltzmann.

How about the appearance of Bose--Einstein statistics in more complex entities of meaning, as the texts produced by human language? Any written text has been considered above as a combination of the concepts corresponding to the words that appear in the text. Again, the illustration of a simple example is useful to grasp the point. Let us consider a written text that contains two instances of the word {\it Cat} in it. It is then trivially clear that, if we exchange in the text the two words {\it Cat}, the meaning content of the text does not change at all. Hence, a written text of human language contains a perfect symmetry for the exchange of identical  words (concepts). This provides an additional intuition that Bose--Einstein statistics should apply whenever the distribution of the words appearing in a given text is calculated.

Relying on the above insights, we have analysed in detail a story-telling text, namely, the Winnie the Pooh story entitled ``In Which Piglet Meets a Haffalump'' \cite{milne1926} and we have calculated the frequency of appearance of all the words in the text. Next, we have interpreted these frequencies of appearance as the cognitive counterpart of single-entity energy levels, hence micro-states, of identical physical entities. 

We have found that these energy levels significantly differ from the Maxwell--Boltzmann distribution in Equation (\ref{maxwellboltzmanndistribution}), whereas they perfectly distribute according to the Bose--Einstein distribution in Equation (\ref{boseeinsteindistribution}). This has allowed us to conclude that a text of human language can be considered in terms of its meaning content as an ideal gas of identical and indistinguishable conceptual entities, which we have called `cognitons' (see Section \ref{intro}), in complete analogy with the case of an ideal gas of bosons \cite{aertsbeltran2020,beltran2021,aertsbeltran2022a,aertsbeltran2022b}. Furthermore, we have provided arguments to support the hypothesis that the gas of cognitons is in an overall state that is close to a Bose--Einstein condensate \cite{aertsbeltran2020}. The analysis has been implemented on various texts, including short and long stories, e.g., novels, and we always found the results that we summarize in the following.

In our theoretical scheme, each word of a text can be considered as a conceptual entity in a specific micro-state whose energy level is defined by the number of times the word appears in the text in question. The analogy that inspires us to do this is the following. A black body radiates photons of different frequencies, and the law of black body radiation shows us the frequency of appearance of a photon of a certain energy level. Similarly, a text telling a story can be considered as an entity emitting words to a possible listener or reader, and the frequencies of appearance of certain words -- words are energy states of cognitons -- will then give the magnitude of the energy according to the prevailing radiation law. More precisely, the most frequent word is given by the lowest energy level $E_0$. Let us suppose that the text contains $n+1$ different words and let us order them according to their increasing energy level or, equivalently, according to their decreasing number of appearance in the text. Let us then set, for a given word $w_i$, its energy $E_i=i$, $i=0,1,\ldots,n$. This means that we set $E_0=0$ as the `ground state energy'. For example, in the Winnie the Pooh story ``In Which Piglet Meets a Haffalump'', the most frequent word is {\it And}, appearing $133$ times, and this word is given an energy $E_0=0$.\footnote{One may wonder why we give the lowest energy level the value zero. There is a long but complex way of doing so in physics in the studies of the radiation law and, more specifically, when Bose--Einstein condensation is studied, hence we use the same calibration here.  For our analysis to become equivalent to the analysis of Zipf's law (see also footnote \ref{zipf}), we would have to choose the value $1$ for the lowest energy level, however, that these are not fundamental differences but just different calibrations is analysed in detail and explained in \cite{aertsbeltran2020}.} Next, we have the word {\it He}, appearing 111 times, and this word is given energy $E_1=1$, and so on. Overall, the story contains 543 different words, thus 543 different energy levels.\footnote{We stress here an important difference between physics and cognition with respect to the measure of energy. In physics, the unit of energy is a derived quantity, whereas energy becomes a fundamental quantity in cognition, where the cognitive equivalent of physical space cannot be uniquely identified.}

Let now $N(E_i)$ be the number of times the word $w_i$, with energy $E_i$, appears in the text. Hence, the `total number of words' appearing in the text is
\begin{equation} \label{totalnumber}
N=\sum_{i=0}^{n} N(E_i)
\end{equation}
For example, in the Winnie the Pooh story ``In Which Piglet Meets a Haffalump'', the total number of words is $N=2,655$. 

It follows from the above that the $N(E_i)$ words $w_i$ in the text are associated with the overall energy $E_i N(E_i)$, $i=0,1,\ldots,n$. For example, in the Winnie the Pooh story ``In Which Piglet Meets a Haffalump'', the energy level $E_{54}=54$ corresponds to the word {\it Thought} which appears a number $N(E_{54})=10$ of times in the story. Thus, the overall energy associated with all the words {\it Thought} appearing in the text is $E_{54}N(E_{54})=54\cdot 10=540$. 

Hence, the `total energy of words' in a story is
\begin{equation} \label{totalenergy}
E=\sum_{i=0}^{n} E_i N(E_i)=\sum_{i=0}^{n} i N(E_i)
\end{equation}
For example, in the Winnie the Pooh story ``In Which Piglet Meets a Haffalump'', the total energy of words is $E=242,891$.

Looking at Equations (\ref{totalnumber}) and (\ref{totalenergy}), we can see that both $N$ and $E$ can be retrieved from empirical data, namely, word counts in this case, exactly as in physics.

\begin{figure}
    \centering
    \subfloat[Numbers of appearances distribution graphs]{{\includegraphics[width=8cm]{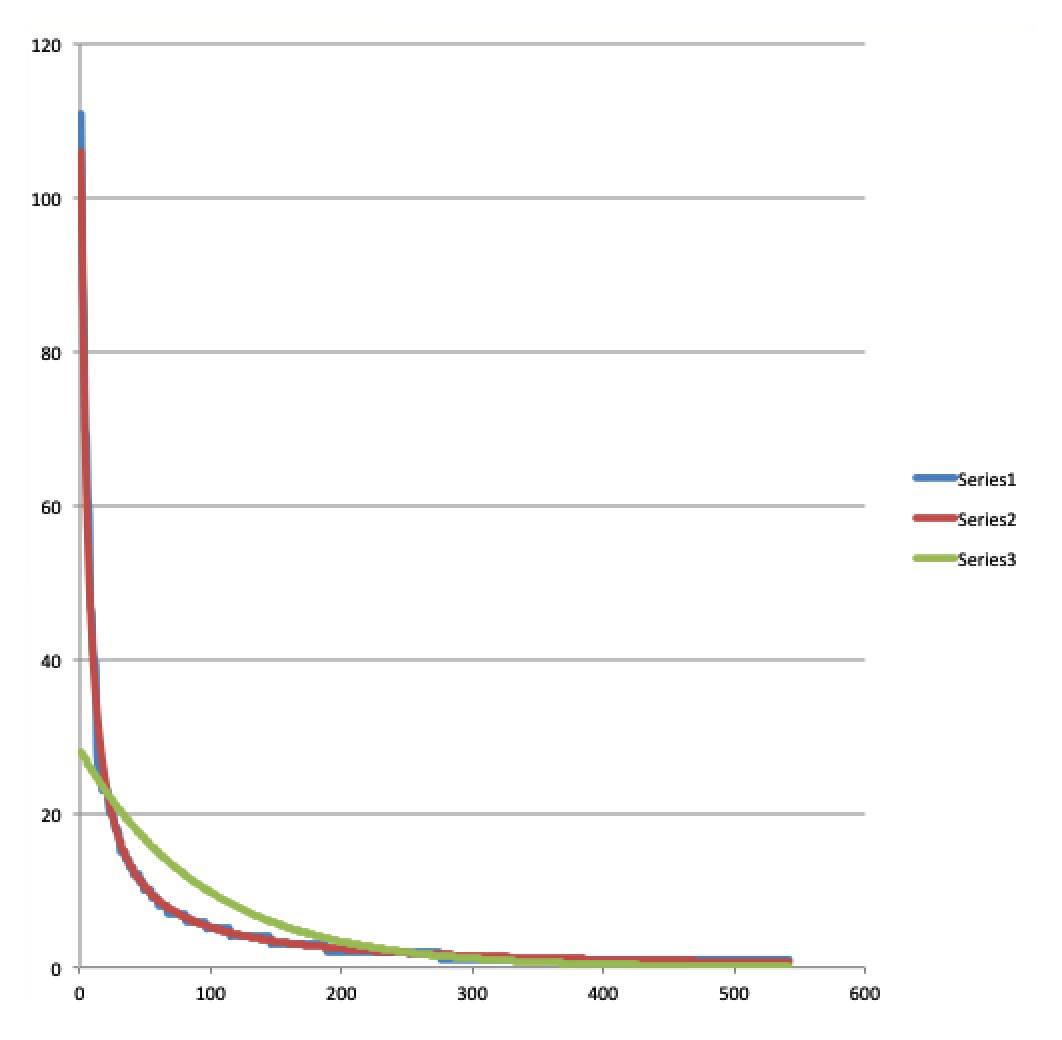} }}%
    \qquad
    \subfloat[
    A $\log/\log$ plot of the numbers of appearances distribution graphs]{{\includegraphics[width=8cm]{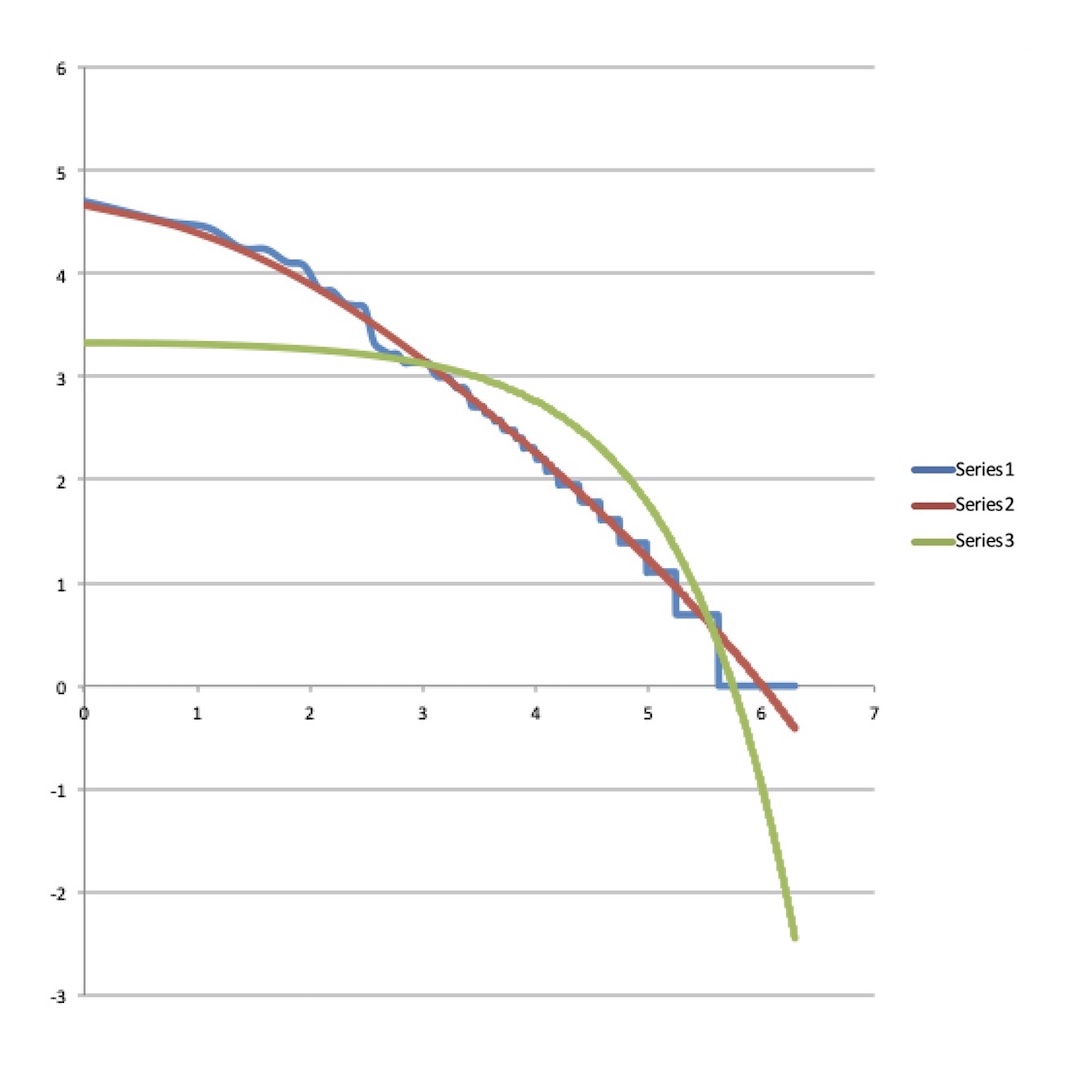} }}%
    \caption{In (a) we represent the number of appearances $N(E_i)$ of words in the Winnie the Pooh story ``In Which Piglet Meets a Haffalump'' 
    \cite{milne1926}, ranked from lowest energy level, corresponding to the most often appearing word, to highest energy level, corresponding to the least often appearing word. The blue graph (Series 1) represents the data, i.e. the collected numbers of appearances from the story, the red graph (Series 2) is a Bose--Einstein distribution model for these numbers of appearances, and the green graph (Series 3) is a Maxwell--Boltzmann distribution model. In (b) we represent the $\log / \log$ graphs of the numbers of appearances and their Bose--Einstein and Maxwell--Boltzmann models. The red and blue graphs coincide almost completely in both plots, whereas the green graph does not coincide at all with the blue graph of the data. This shows that the Bose--Einstein statistical distribution is a good model for the numbers of appearances, while the Maxwell--Boltzmann distribution is not.}%
    \label{piglethaffalunmpgraphpiglethaffalunmploggraph}%
\end{figure}

\begin{figure}
\centering
\includegraphics[width=8cm]{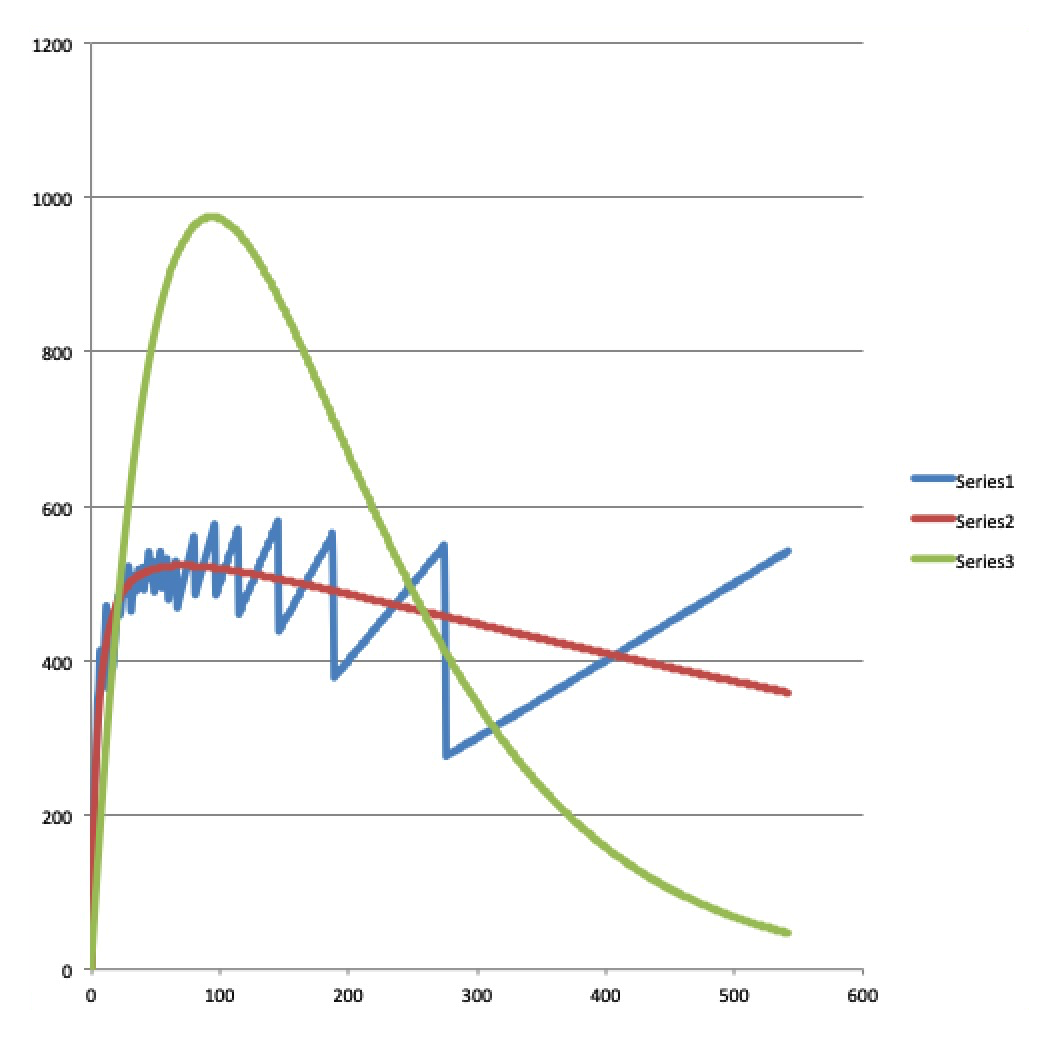}
\caption{The figure reports the energy distribution of the Winnie the Pooh story ``In Which Piglet Meets a Haffalump'' 
\cite{milne1926}. The blue graph (Series 1) represents the energy $E_iN(E_i)$ radiated by the story per energy level $E_i=i$, 
the red graph (Series~2) represents that same energy  per energy level as modelled by the Bose--Einstein distribution, and same for the green graph (Series 3), which follows the Maxwell--Boltzmann distribution.}
\label{piglethaffalunmpenergygraph}
\end{figure}

In our theoretical scheme, the numbers of appearance $N(E_i)$, $i=0,1,\ldots, n$, in Equations (\ref{totalnumber}) and (\ref{totalenergy}) can be used to determine the pairs of constants $(A,B)$ and $(C,D)$ in Equation (\ref{boseeinsteindistribution}) and (\ref{maxwellboltzmanndistribution}), respectively, subject to the constants $N=2,655$ and $E=242,891$, thus checking whether the Bose--Einstein distribution or the Maxwell--Boltzmann distribution fit the data. This task has effectively been accomplished in  \cite{aertsbeltran2020,aertsbeltran2022a,aertsbeltran2022b}, comparing the ensuing Bose--Einstein and Maxwell--Boltzmann distributions with empirical data. Figures \ref{piglethaffalunmpgraphpiglethaffalunmploggraph} and \ref{piglethaffalunmpenergygraph} summarise the comparison. 

The results of the comparison significantly reveal that `Bose--Einstein statistics is in remarkably good fit with empirical data', whereas a `large deviation is observed in the data from Maxwell--Boltzmann statistics'. The conclusion is that, once a notion of energy is adequately introduced and quantified from empirical data, a collection of words in a written text produced by human language behaves as a suitable gas of identical bosons. As a matter of fact, we have introduced the term `cogniton' 
as the fundamental quantum of cognition. Equivalently, each word $w_i$ with energy $E_i$ corresponds to a micro-state of a cogniton with energy $E_i$. The overall text then behaves as a `gas of cognitons' whose energies are distributed according to Bose--Einstein statistics.\footnote{It is worth to mention, at this stage, that we have also provided a theoretical foundation of `Zipf's law' in human language \cite{zipf1935,zipf1949}, whose theoretical origin is not understood yet and whose nature is therefore considered to be merely empirical. On the contrary, our results in \cite{aertsbeltran2020,aertsbeltran2022a,aertsbeltran2022b} derive it on the basis of the presence of a Bose--Einstein behaviour in the energy levels of words in a text. \label{zipf}}

Let us finally once again take up the example of the farmer and the child in Section \ref{example}, which is a cognitive situation where, when the farmer makes the choices in the way we have explained, a Maxwell--Boltzmann distribution results while, if the child makes the choices, a different from Maxwell--Boltzmann distribution will likely follow, possibly a Bose--Einstein one. Why this difference in the statistical distributions? To answer this question, we can note the following difference. The farmer chooses using the contextual environment of the physical bodies of the cats and dogs and the physical objects that are the kennels in which they are placed. All the physical situations that pass through the dynamics of the farmer's choice mechanism are distinguishable and even follow a readily identifiable trajectory in time and space. Furthermore, no dependence is introduced between the first and second choice phases, and Maxwell--Boltzmann statistics results as a consequence of this independence. The child, on the other hand, chooses primarily between three situations in which either two puppies, or two kittens, or a puppy and a kitten will end up at home with her or him to play with. Hence, for the choice dynamics in which the child is contextually immersed, the difference between a kitten first and a puppy second, or a puppy first and a kitten second, which is essential in the Maxwell--Boltzmann case to assign to it a double probability, namely $1/2$, compared to the probability assigned to two kittens, namely $1/4$, or that assigned to two puppies, likewise $1/4$, is totally absent and irrelevant. After all, when both pets play with each other and with the child, it is totally unimportant which of the two was chosen first and which second. The consequence of that is that puppy and kitten do `not' figure as independent variables in the dynamics of the child's contextual choice. 

It would of course be possible to test the specific example we have considered using the calibrated method in psychology, with a 
suitably 
chosen Likert scale, then make an experimental estimate of the statistics with which children would choose between two kittens, two puppies or a kitten and a puppy. This specific experiment was however not carried out by us, but we examined very similar experimental situations, which also included a choice between cats and dogs, being able to show that the Bose--Einstein statistics systematically constitutes a better model than the Maxwell--Boltzmann statistics \cite{aertssozzoveloz2015,aertsbeltran2020}. 

The phenomenon we have called `contextual updating' contains a possible explanation for this statistical behaviour. In fact, the conceptual combinations  `two kittens', `two puppies', and `a kitten and a puppy',  each behave differently with respect to the entire context of meaning that characterizes a child's choosing, and in this sense they acquire their own individual probabilities of occurrence. Does this mean that the probabilities of these choices are necessarily equal and that a Bose--Einstein statistics follows? Not necessarily, contextual updating will only cause them to be associated with non-decomposable probabilities, thus failing the well-known property, typical of the Maxwell--Boltzmann statistics, that the combination `one kitten and one puppy' possesses twice the frequency of `two kittens' and `two puppies'. In other words, the statistical independence of the choices `kitten' and `puppy' is lost as a consequence of a mechanism, that of contextual updating, that we identified as lying also at the origin of the violation of Bell's inequalities \cite{ijtp2023}. 

Although we will not elaborate on this in this article, we would like to point out that the fact that Bose--Einstein statistics prescribes the same probabilities of occurrence for the different conceptual combinations is perhaps due to a statistical averaging effect, similar to the `universal average' that we introduced and studied in relation to Born probabilities \cite{aertssassoli2014,aertssassoli2015}.  Indeed, if the contextual updating mechanism underlies the loss of statistical independence, we can imagine that each of the possible combinations could elicit a multitude of different meanings. In our example, a child choosing may be influenced by the fact that adult cats and dogs are often not friends with each other, making the choice  `a kitten and a puppy' less attractive than the choice `two kittens' or `two puppies'. However, there are also contexts of meaning that make this choice more attractive, for example, when a child wants to play both with a kitten and a puppy.  This multitude of meaning contexts correspond to `different ways of choosing' the available conceptual combinations, and each way of choosing is associated with a different statistics. When an average over them is considered, this is expected to lead to a Bose--Einstein type of quantum statistics  \cite{aertssassoli2014,aertssassoli2015}.

Note that although the contextual updating mechanism is related to `meaning', in the analysis of our example the presence of the Bose--Einstein statistics is not necessarily related to a dynamics caused by the presence of meaning, it being sufficient that the context in which the choice process takes place is connected with a spatiotemporal structure. In other words, for the time being we must remain cautious when highlighting the possible role of meaning dynamics in physical systems, since a non-spatiotemporal context does not necessarily correspond to a context of a cognitive-conceptual nature, although the latter is certainly an emblematic example of a reality not entirely ascribable to a spatiotemporal theatre. We will return to this point in more detail in  Section \ref{temperature}.

\section{Randomness and temperature\label{temperature}}
We believe that the way in which identity and indistinguishability have been identified in human language may provide new powerful insights into how identity and indistinguishability should be looked at in QM as a theory of physical systems. We have seen in the farmer example in Section \ref{example} that different words are not indistinguishable and, more, it is with such not indistinguishable words that Bose--Einstein statistics arises, by thoroughly associating with them probabilities of occurrence that cannot come from independent choices for the individual cases, namely, the probability of occurrence of different words in a text are not independent from each other. It was exactly this lack of statistical independence that Einstein found so disturbing in the case of photons. He saw no other possibility than concluding that the probabilities of occupation of a given micro-state are not independent for the individual photons.  As we have mentioned in Sections \ref{intro} and \ref{early}, Einstein attributed this lack of statistical independence to a mysterious force among the photons. 

We intend to explore in this section the role that `meaning' might play in this lack of statistical independence. Let us support our analysis with an example and consider again the Winnie the Pooh story ``In which Piglet meets a Haffalump'' which we have discussed in Section \ref{language}. The word {\it Piglet} occurs $47$ times in the story, while the word {\it First} occurs $7$ times. Let us now suppose that one would have asked the author to write some additional paragraphs to the story text. One realizes at once that the probability that the word {\it Piglet} would occur in these additional paragraphs is higher than the probability that the word {\it First} appears in them. The reason is that the story text carries an overall meaning, hence the words that have more affinity with such overall meaning have a higher chance of appearing when the story is continued to be written. Equivalently, one can say that `meaning is a force that makes the same words attract each other to clump together as a result of this meaning force' \cite{aertsbeltran2022b}. Hence, meaning could be what plays the conceptual counterpart of the force that Einstein believed to cause photons to clump into the same state. Each word that is added to the text is connected with all the words already existing in the text if the latter is telling a story that possesses coherence in terms of meaning, as is usually the case for stories told by human beings. 
 
However, what if we consider a text where we have removed some of the meaning the text originally contained? Let us explain what we mean and illustrate this concretely using the Winnie the Pooh story, which contains a total of $2,655$ words of which there are $543$ different words.

Let us now create a new collection of $2,655$ words where each word is chosen at random from the $543$ different words that were 
originally used for the Winnie the Pooh story. It is not that we have removed all meaning in this way because, after all, the words themselves carry meaning on their own and it is still the words of the Winnie the Pooh story that appear in the collection of words chosen at random in this way. Yet, we have negated the extra meaning that is precisely this unique combination that makes them the Winnie the Pooh story when placed in their original sequence. And we have introduced randomness that does allow each word to take its place independently of the other words that are there. This new collection of words cannot be considered as telling a story, but it has the same length as the Winnie the Pooh story, namely $2,655$ words, and is made up of exactly the same words, though differently arranged. When we set up our typical thermodynamic radiation scheme with this collection of words, we can see that the largest frequency of the same word appearing is $13$ or $14$, while it was $133$ for the Winnie the Pooh story. These numbers, $13$ or $14$, showed up for the dozen or so random choices we tried out by using the tools offered on the website \url{https://www.random.org}, where a true random choice is guaranteed. Admittedly, there will be fluctuations where more of the same words show up than the $13$ or $14$ that occurred in our tries, but the chances become very small that randomly many more words turn out to be the same, let alone $133$, as in the Winnie the Pooh story. 

The question we ask first and foremost is whether this collection of words randomly chosen from the collection of words that make up the Winnie the Pooh story can still be modelled with the Bose--Einstein distribution as was the case for the original Winnie the Pooh story, and more, whether the Maxwell--Boltzmann distribution still does not provide a good model at all. 
\begin{figure}
\centering
\includegraphics[width=8cm]{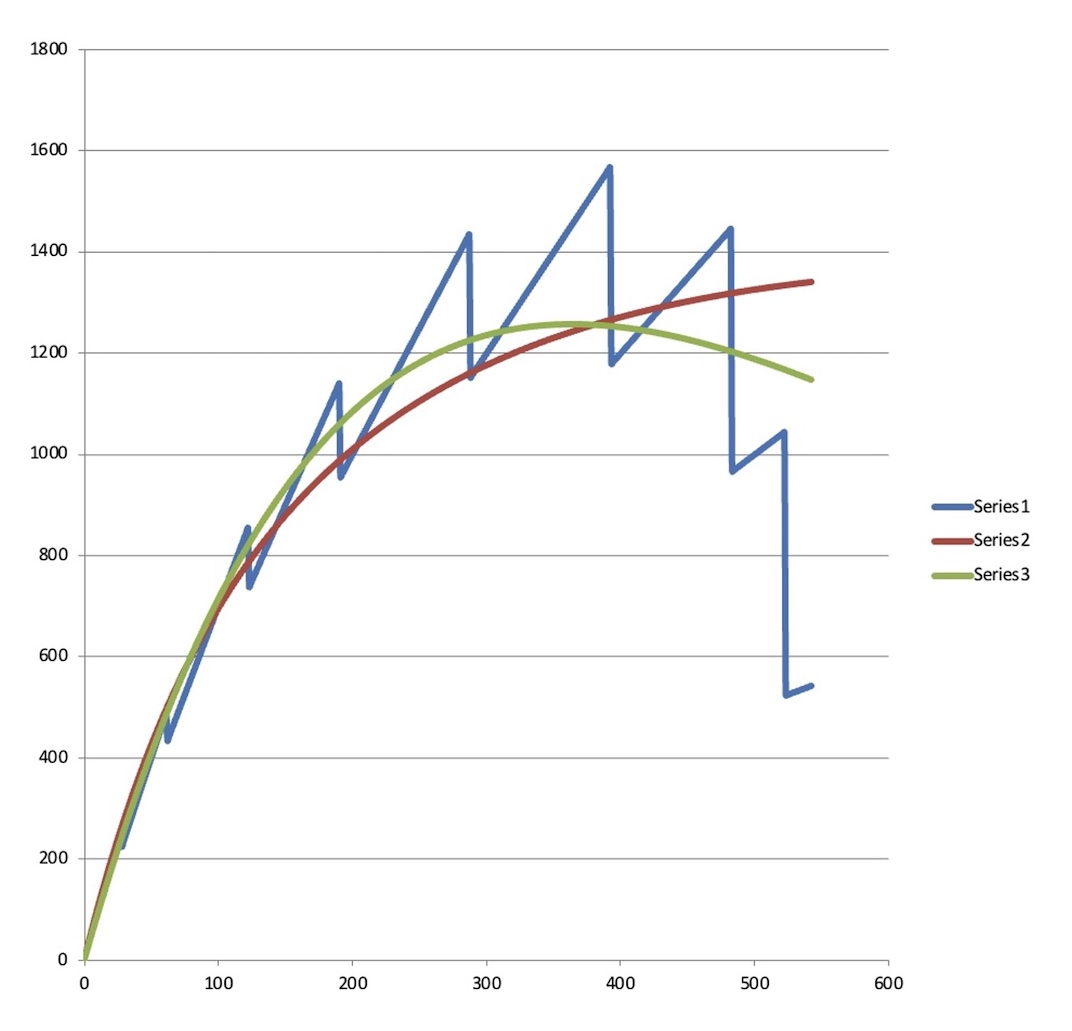}
\caption{The figure reports the energy distribution of the randomized Winnie the Pooh story. The blue graph (Series 1) represents the energy radiated per energy level, the red graph (Series 2) represents the energy radiated by the Bose--Einstein model per energy level, and the green graph (Series 3) represents the energy radiated by the Maxwell--Boltzmann model per energy level.}
\label{randomaddedenergygraph}
\end{figure}
\begin{figure}
\centering
\includegraphics[width=8cm]{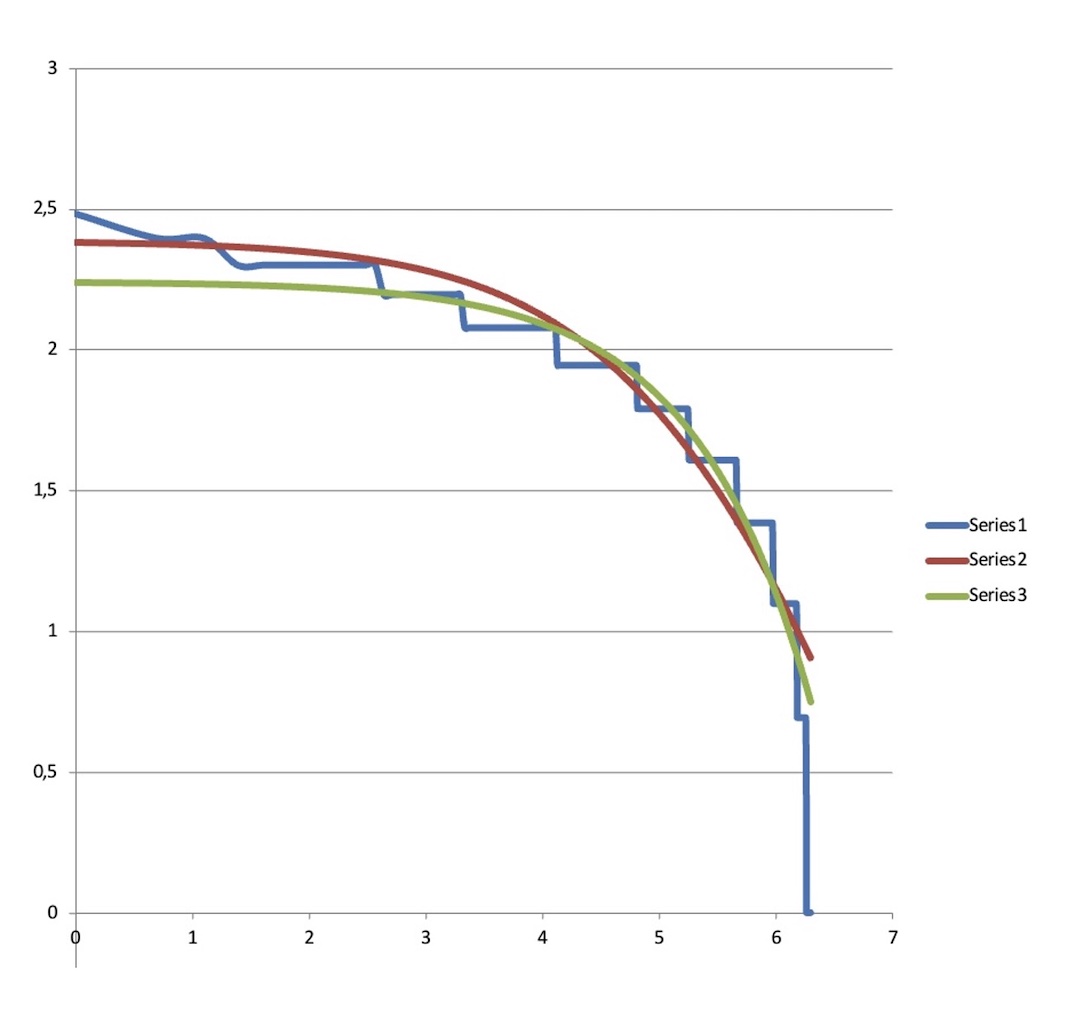}
\caption{The figure reports the $\log / \log$ plot of the collected numbers of appearances and their Bose--Einstein and Maxwell--Boltzmann models. The  red and blue graphs coincide almost completely, whereas the green graph does not coincide with the blue graph of the data.}
\label{randomaddedloggraph}
\end{figure}
The graphs in Figures \ref{randomaddedenergygraph} and \ref{randomaddedloggraph} give us an answer to our questions. It is clear that even the collection of words that occurs in the randomised Winnie the Pooh story can be better modelled by a Bose--Einstein distribution than it is the case by a Maxwell--Boltzmann distribution, although the difference between the two is much smaller than it is for the original Winnie the Pooh story, the slight superiority of the Bose--Einstein distribution being able to be seen clearly only in the $\log / \log$ plot of Figure  \ref{randomaddedloggraph}. Thus, even if we introduce complete randomness in choosing which word from the collection of the $543$ 
different words of the Winnie the Pooh story, to be placed in the $2,655$ places corresponding to the total number of words of the story,  
the Bose--Einstein distribution still gives a good fit with the experimental data, but also the Maxwell--Boltzmann is much less off this time than it was with the data from the unchanged original story (see Figures \ref{piglethaffalunmpgraphpiglethaffalunmploggraph} (b) and \ref{piglethaffalunmpenergygraph}, Section \ref{language}). 

The important new element to understand what is happening is the increase in total energy, which was equal to $242,891$ in the case of the original unchanged Winnie the Pooh story, and now becomes equal to $546,430$, that is, more than doubling, in the case of the Winnie the Pooh words randomly placed in the available places of the story. More interesting still, however, is comparing the constant $B$ in Equation (\ref{boseeinsteindistribution}) and the values it takes on in the two cases. For the unchanged Winnie the Pooh story we have $B=593.5$, while for the randomized version of the Winnie the Pooh words we have $B = 2152.7$, which is four times larger. This constant $B$ is the temperature expressed in energy units, or, more fundamentally, it is the energy that is not usable to do work with, i.e. heat energy. 
In other words, randomizing the Winnie the Pooh words has increased the heat of the text by a factor of four. 

As a final remark, we note that, if we take both formulas, the one describing the Bose--Einstein distribution in Equation (\ref{boseeinsteindistribution}), and the one describing the Maxwell--Boltzmann distribution in Equation (\ref{maxwellboltzmanndistribution}), and remember that the constant in the exponential represents the temperature or, better, the form of energy we call `heat', then the exponential becomes equal to $1$ for the temperature going to infinity, and both formulas then become a constant. In other words, for infinite heat, all energy levels radiate equally, and the Bose--Einstein distribution and the Maxwell--Boltzmann distribution coincide, and are constant. The `number of appearances distribution graph' then becomes a horizontal straight line. This is, of course, a theoretical limit case that does not occur in reality, but it does make us understand why, as the temperature increases, the slope of the interpolation line of the $\log/\log$ graph changes and tends to zero (the line rotates counterclockwise) in the limit of an infinite temperature, that is, in the limit where we go from an original story, where no heat has been added yet and therefore is optimally close to the Bose-Einstein condensate situation in terms of the presence of `meaning’ (see Section \ref{hypothesis} for more explanation), to a situation where heat energy is increasingly added by increasing randomness in the choice of words.

We repeated this analysis of introducing an element of randomness, consisting of spreading the numbers of appearance of the words originally used in a story over its placeholders, for the other stories we studied in \cite{aertsbeltran2020}, entitled ``The Magic Shop'' and ``Gulliver's Travels''. And we obtained even more data for this analysis of heat and temperature by implementing this randomness as well on several other stories of novels we studied using corpora of the Italian language \cite{italiancorpora2024}. The result was the same, namely, the total energy and temperature of each of the stories increased in a way that is similar to that of the Winnie the Pooh story. We do not give here details of our analysis of the thermodynamic temperature of stories in human language and refer to \cite{italiancorpora2024,temperature2024} for the complete study, but draw the following conclusion as to its importance for the subject of our article.\footnote{It is remarkable that, despite the specific linguistic differences between English and Italian languages, we have identified in the study of Italian novels a statistical behaviour that is completely analogous to what we have found in English texts, with respect to both Bose--Einstein distribution and appearance of effects due to temperature in the presence of randomization. This preliminary analysis shows that there are general meaning-related patterns in human language that are `species-specific'. Hence, we expect that similar patterns can also be identified if different languages are analysed.}
Even when a specific part of the meaning contained in a story is taken away and the words are placed at random in a new text, a Bose--Einstein statistical model still exists for it, which means that even these more random collections of words carry the phenomenon of condensation, that is, the clumping together of the same words.

One might think that a text generated from the Winnie the Pooh story, by randomly choosing words from those available in the original writing, would be stripped of all meaning. However, we can assume that a residual level of meaning is present also in a text `heated up' in this way. Indeed, not all the words available in the English language participate in the random selection. This forces the different energy levels of the gas of cognitons to be more densely populated, as we need to have some degree of repetition of the used words, and this repetitiveness produces a possible perception of greater meaning than for a text where almost every randomly chosen word would be different (as it would be the case if all words in the English language participated in the random selection process, the order of magnitude of their number being $200,000$). 

However, it should be added that we have introduced randomness into the original story by using the already existing fabric of words that form it, and this can be interpreted as implementing a certain type of locality. In fact, placeholders for words in the original story can be regarded as precursors of a `locality claimed by words'. Note in this regard that when we index the placeholders, those words that would otherwise be absolutely indistinguishable from each other actually become distinguishable (for example, to the fourth time that `Piglet' appears in the story of Winnie the Pooh, we can assign index 4). In this sense, the specific situation in which only the words of the original story are subjected to randomisation may be similar to the situation of a gas that has been entirely localised in a container in a given physics laboratory. Indeed, molecules of the same gas located in another container, located in another laboratory, will play no role in determining its temperature. If we consider experiments in which a Bose--Einstein condensate is made from a specific gas, and let the temperature slowly rise again, we will go through a phase in which quantum coherence is indeed still partially present, just as in the randomisation we introduced `meaning' is still partially present. In this regard, it is interesting to ask what phase a star like our sun is currently in \cite{temperature2024}. 

\section{The possible role of meaning \label{hypothesis}}

Let us first make it clear that we do not consider here all aspects of how meaning plays a role in the dynamics that takes place and in the structure that is present in human language, also because several of these aspects are currently the subject of ongoing research. However, already at this stage, we can draw from our analysis in Sections \ref{language} and \ref{temperature} some interesting explanatory hypothesis and insights. There is a specific mechanism that we identified in one of our previous publications \cite{ijtp2023}, which we called `contextual updating', that incorporates an important part of what meaning does. 

More precisely, we have proved in several articles that conceptual meaning can be represented by entanglement in the Hilbert space formalism of QM (see, e.g.,  \cite{aertsgaborasozzo2013,aertssozzo2016,aertssassolisozzo2016,pisanosozzo2020,aertssassolisozzoveloz2021,aertsbeltran2022a,philtransa2023,ijtp2023}). Furthermore, we have recently suggested that, in both language and physics domains, entanglement can be considered as a `phenomenon of contextual updating' \cite{ijtp2023,philtransa2023}. According to this view, in both language and physics domains, the lack of statistical independence would be produced by a phenomenon of entanglement through contextual updating. 

To explain what we mean by this mechanism of contextual updating, let us consider the combination of two words in a text. Each word carries meaning but, whenever the two words are combined, their combination also carries its proper meaning, which is not the trivial combination of the meanings associated with the two individual words as prescribed by a classical logical semantics. The new emergent meaning of the combined word arises in a complex contextual way, in which the whole of the context relevant to the story plays a fundamental role. Thus, each time a word is added to a text, a mechanism of updating influenced by the meaning of the whole context occurs, and this updating continues to take place until the end of the story that contains all the words. We proved that this mechanism of contextual updating through which meaning is attributed has to be carried by an entangled state \cite{ijtp2023}. For those familiar with the Hilbert space formalism of QM, and the ensuing tensor product for the description of composite quantum entities, they will recognize such a contextual updating mechanism in the mathematical procedure of describing multiple composite quantum entities. Indeed, whenever a Hilbert  space of an individual quantum entity is coupled to other state spaces via the tensor product, new states form as a consequence of the superposition principle, which always contains a majority of entangled states. It is these superposition states that accomplish the contextual updating in the mathematical  formalism.\footnote{For a detailed investigation of the problems of separability and entanglement in both quantum physics and human language within axiomatic foundations of quantum mechanics, see also \cite{aertsetalPhilTransA2024}.}

We put forward the hypothesis that a similar mechanism of contextual updating occurs in the micro-physical entities described by QM. Indeed, whenever an electron is added to an atom, one has to account for all the other electrons that are already present in the atom, as well as its nucleus, through mutual interactions. The new electron gets entangled with the others. Analogously, whenever a photon is added to a gas of photons, the former gets entangled with all other photons and this occurs through the above mentioned mechanism of contextual updating. 

This mechanism of contextual updating is very characteristic of how `meaning attaches and interweaves itself with' a collection of words. In general, the writer's intention is to make the text as clear as possible, which in our thermodynamic formalism translates 
into a `pure state', with von Neumann entropy equal to zero. Whether this pure state is also pure for the reader and/or listener of the text is an interesting question in itself about which, however, we will not elaborate in this article \cite{temperature2024}. 

To arrive at a formulation of what the role of meaning is we can put forward the following hypothesis. Bose--Einstein remains the dominant statistics even if randomness is introduced artificially, thus we can say that, in general, the phenomenon of clustering of equal states is always present. Meaning, and the extra force-like mechanism 
it brings into being, however, adds something, namely, a striving to realize a pure state for the considered entity, and hence the Bose--Einstein effect is magnified by it, with the Bose--Einstein condensate as a limit case, where only one energy state is left. Regarding the latter, it is not frequent that in the realm of human cognition texts consisting of 
`a collection of words consisting of one and the same word’ are relevant, but there are examples. The text
`stop, stop, stop, \ldots, stop',  i.e. a repetition of the word `stop'  shouted by a parent who sees her or his child making inattentive approaches to cross a busy street, is such an example of a Bose-Einstein condensate, with all cognitons in one energy state, forming a text that is contextually closed in terms of meaning. Of course, such a text will probably be followed by a second text, like for example `wait till I get there', less loud, but still called out by the parent, thus populating more energy levels of the gas of cognitons. 

Hence, in general, meaning, as used by people in human language, attaches and interweaves with states that are close to this Bose--Einstein condensate but usually not equal to it. It is an intriguing, but not yet experimentally solved question, whether things happen similarly in the micro-world in relation to `quantum coherence', which would play an equivalent role to the notion of meaning in the human cognitive domain  \cite{temperature2024}. In any case, we believe that the analysis we have put forward in the present article comes closer to an explanation than the explanation produced by the symmetrization postulate and the ensuing complete indistinguishability of identical quantum physical entities.

To conclude, we have observed that the classical statistical interpretation of thermodynamics rests on the assumption that physical entities behave in a random and independent way at a microscopic level, which entails maximization of entropy \cite{huang1987}. Based on our analysis of human language, we have suggested that it is more appropriate instead to consider that, also in the case of physics, there is no situation where entities are truly random and independent, as (conceptual or physical) entities always bundle (entangle) together, as evidenced by the presence of the Bose--Einstein statistics, with the aim to collaborate and reduce the overall uncertainty, hence entropy, and produce globally a pure state  \cite{ijtp2023,philtransa2023}. Let us use this insight of the collaborative dynamics of words forming a story to complete our critique of the modern textbook introduction to quantum statistics. Indeed, we believe that this collaborative dynamics, that is, that words in a story work together and form their individual state accordingly, so that the whole story would be optimal in terms of the meaning the human mind is able to grasp in it, most fully defines the nature of words as quanta of human language. Thus, what follows is to be seen as a complement to what we have already presented in the previous pages, and in the articles cited above, considering in particular the issue of lack of independence, which has been of such burning concern in the early decades of QM. 

Consider, e.g., the sentence {\it Eleven Horses}. We have already noted that in this sentence each of the horses is totally indistinguishable from the others. But is this statement really what is most important to characterize what is going on with each of the eleven horses in this sentence? It is much more relevant to note that {\it Horse} is an abstraction that refers to concrete horses, which can be of different types, with bodies located in space. But, still more relevant, is perhaps that {\it Horse} is a concept and not an object. Even more, suppose the above sentence is part of a somewhat longer sentence, namely the following: {\it We Encountered Eleven Horses On Our Walk}. Then, we see that the piece of sentence {\it Eleven Horses} changes by a mechanism we have called `contextual updating', and now suddenly deals with horses in the neighbourhood where we were walking. But the mechanism of contextual updating is but one of the fantastically complex tools of subtle changes which words, the quanta of human language, are capable of. The most comprehensive 
way to grasp this complex adaptive dynamics of words is to state that `words collaborate and are willing to become more uncertain (more abstract), with the intention of making the story they are part of as certain (concrete) as possible'. And it is this intention, to minimize 
the entropy of the story the words are part of, that makes them choose the state in which they offer themselves. That a story consists of words some of which are identical, although they probably take on different roles in how they contribute to lowering the entropy of the whole story, considering also where in the story they appear, makes it possible for these identical words to be interchanged without changing the story, which is the `exchange symmetry' present in human language. 

We can describe, however, even more in depth the way in which words collaborate to build a sentence or a story, taking into account that in human language such collaboration aims to come as close as possible to carrying the meaning that the person wishes to communicate with 
that specific story. The mind of the person uttering it attempts to choose the words that do so in the most appropriate way. Would it make sense to try to grasp this subtle dynamics by assuming that words are completely indistinguishable? Clearly not. In the sentence {\it Eleven Horses}, it can be meaningfully asserted that the horses are indistinguishable, but the reason is the ``sacrifice'' that 
this sentence makes to the story in which it appears, with the intention of making that story what it is desired to be by the person uttering it. This happens because of the ``force'' exerted by the meaning to which the collaborative dynamic of words submits.
In a structurally very similar way, a collaborative dynamics of molecules or photons submits to the ``force'' exerted by the quantum coherence of the domain they are part of. 

Concerning in particular the choice of assigning equal probability to each micro-state, which is an aspect usually not explained in the standard formulation of modern textbooks, we believe that it is related to the contextual updating mechanism that submits each micro-state to the overall meaning of the story, or to the entire quantum coherence domain of the gas in question. We do not rule out the possibility that this choice is an expression of a universal averaging over different probabilities of occurrence, for each event. A refinement of experiments, in the future, might shed light on these subtle dynamics of molecules, or photons, and highlight some of the mechanisms suggested above, which we know are present in the case of human language.

Our results clearly indicate that a theoretical direction towards the development of a quantum-type thermodynamics that is alternative to existing formal approaches to `quantum thermodynamics for physics', none of which is satisfying  \cite{gemmeretal2009,mahler2015,gogolineisert2016}, is a valuable aim to pursue. A possible direction for such an alternative theoretical approach rests on the notions of energy and entropy as defined in cognitive-linguistics domains, and we refer to  \cite{philtransa2023} for more details about the development of a thermodynamics of human cognition and human culture.

\section*{Acknowledgements}
This work was supported by the project ``New Methodologies for Information Access and Retrieval with Applications to the Digital Humanities'', scientist in charge S. Sozzo, financed within the fund ``DIUM -- Department of Excellence 2023--27'' and by the funds
that remained 
at the Vrije Universiteit Brussel 
at the completion of 
the ``QUARTZ (Quantum Information Access and Retrieval Theory)'' project, part of the ``Marie Sklodowska-Curie Innovative Training Network 721321'' of the ``European Unions Horizon 2020'' research and innovation program, with Diederik Aerts as principle investigator for the Brussels part of the network.


\begin{thebibliography}{}

\bibitem{huang1987} Huang, K. (1987). {\it Statistical Mechanics}. New York: Wiley.

\bibitem{klein1961} Klein, M. J. (1961). Max Planck and the beginnings of the quantum theory. {\it Archive for History of Exact Sciences 1}, 459--479.

\bibitem{howard1990} Howard, D. (1990). ``Nicht Sein Kann was Nicht Sein Darf," or the prehistory of EPR, 1909--1935: Einstein's early worries about the quantum mechanics of composite systems. In A. I. Miller (Ed.), {\it Sixty-Two Years of Uncertainty}, pp. 61--111. Boston: Springer.

\bibitem{darrigol1991} Darrigol, O. (1991). Statistics and Combinatorics in Early Quantum Theory I \& II. {\it Historical Studies in the Physical and Biological Sciences 21}, 17--79 \& 237--298.

\bibitem{monaldi2009} Monaldi, D. (2009). A note on the prehistory of indistinguishable particles. {\it Studies in History and Philosophy of Modern Physics 40}, 383--394.

\bibitem{perezsauer2010} P\'erez, E. and Sauer, T. (2010). Einstein's quantum theory of the monatomic ideal gas: Non-statistical arguments for a new statistics. {\it Archive for History of Exact Sciences 64}, 561--612. 

\bibitem{gorroochurn2018} Gorroochurn, P. (2018). The end of statistical independence: The story of Bose--Einstein Statistics. {\it The Mathematical Intelligencer 40}, 12--17.

\bibitem{einstein1905} Einstein, A. (1905). \"Uber einen die Erzeugung und Verwandlung des Lichtes betreffenden heuristischen Gesichtspunkt. {\it Annalen der Physik 17}, 132--148.

\bibitem{planck1900} Planck, M. (1900). Zur Theorie des Gesetzes der Energieverteilung im Normalspectrum. {\it Verhandlungen der Deutschen Physikalischen Gesellschaft 2}, 237--245.

\bibitem{ehrenfest1911} Ehrenfest, P. (1911). Welche Z\"uge der Lichtquantenhypothese spielen in der Theorie der W\"armestrahlung eine wesentliche Rolle?. Annalen der Physik 341, 91--118.

\bibitem{bose1924} Bose, S. N. (1924). Plancks Gesetz und Lichtquantenhypothese. {\it Zeitschrift f\"ur Physik 26}, 178--181.

\bibitem{einstein1924} Einstein, A. (1924). Quantentheorie des einatomigen idealen Gases. {\it Sitzungsberichte der Preussischen Akademie der Wissenschaften, Physikalisch-mathematische Klasse}, 261--267. 

\bibitem{einstein1925a} Einstein, A. (1925a). Quantentheorie des einatomigen idealen Gases 2. {\it Abhandlung. Sitzungsberichte der Preussischen Akademie der Wissenschaften, Physikalisch-mathematische Klasse}, 3--14. 

\bibitem{einstein1925b} Einstein, A. (1925b). Zur Quantentheorie des idealen Gases. {\it Abhandlung. Sitzungsberichte der Preussischen Akademie der Wissenschaften, Physikalisch-mathematische Klasse}, 18--25.

\bibitem{schrodinger1926a} Schr\"odinger, E. (1926a). Zur Einsteinschen Gastheorie. {\it Physikalische Zeitschrift 27}, 95--101.

\bibitem{dieks2023} Dieks, D. (2023). Emergence and identity of quantum particles. {\it Philosophical Transactions of the Royal Society A 381}, 20220107.

\bibitem{aertssozzo2016} Aerts, D. and Sozzo, S. (2016). Quantum structure in cognition: Origins, developments, successes and expectations. In Haven, E. and Khrennikov, A.Y. (Eds.), {\it The Palgrave Handbook of Quantum Models in Social Science: Applications and Grand Challenges}, pp. 157--193. Palgrave \& Macmillan: London.

\bibitem{aertssassolisozzo2016} Aerts, D., Sassoli de Bianchi, M. and Sozzo, S. (2016). On the foundations of the Brussels operational-realistic approach to cognition. {\it Frontiers in Physics 4}, 17.

\bibitem{pisanosozzo2020} Pisano, R. and Sozzo, S. (2020). A unified theory of human judgements and decision-making under uncertainty. {\it Entropy 22}, 738.

\bibitem{aertssassolisozzoveloz2021} Aerts, D., Sassoli de Bianchi, M., Sozzo, S. and Veloz, T. (2021). Modeling human decision-making: An overview of the Brussels quantum approach. {\it Foundations of Science 26}, 27--54.


\bibitem{aertsaerts1995} Aerts, D. and Aerts, S. (1995). Applications of quantum statistics in psychological studies of decision processes. {\it Foundations of Science 1}, 85--97.

\bibitem{vanrijsbergen2004} van Rijsbergen, C. J. (2004). {\it The Geometry of Information Retrieval}. Cambridge: Cambridge University Press.

\bibitem{aerts2009a} Aerts, D. (2009a). Quantum structure in cognition. {\it Journal of Mathematical Psychology 53}, 314--348.

\bibitem{pothosbusemeyer2009} Pothos, E. and Busemeyer, J. (2009). A quantum probability explanation for violations of `rational' decision theory. {\it Proceedings of the Royal Society B 276}, 2171--2178. 

\bibitem{khrennikov2010} Khrennikov, A. Y. (2010). {\it Ubiquitous Quantum Structure}. Berlin: Springer.

\bibitem{busemeyerbruza2012} Busemeyer, J. and Bruza, P. (2012). {\it Quantum Models of Cognition and Decision}. Cambridge: Cambridge University Press.

\bibitem{aertsbroekaertgaborasozzo2013} Aerts, D., Broekaert, J., Gabora, L. and Sozzo, S. (2013). Quantum structure and human thought. {\it Behavioral and Brain Sciences 36}, 274--276. 

\bibitem{aertsgaborasozzo2013} Aerts, D., Gabora, L. and Sozzo, S. (2013). Concepts and their dynamics: a quantum theoretic modeling of human thought. {\it Topics in Cognitive Science 5}, 737--772.

\bibitem{havenkhrennikov2013} Haven, E. \& Khrennikov, A. Y. (2013). \emph{Quantum Social Science}, Cambridge: Cambridge University Press.

\bibitem{kwampleskacbusemeyer2015} Kvam, P., Pleskac, T., Yu, S. and Busemeyer, J. (2015). Interference effects of choice on confidence. {Proceedings of the National Academy of Science of the USA 112}, 10645--10650. 

\bibitem{dallachiaragiuntininegri2015a} Dalla Chiara, M. L., Giuntini, R. and Negri, E. (2015). A quantum approach to vagueness and to the semantics of music. {\it International Journal of Theoretical Physics 54}, 4546--4556. 

\bibitem{dallachiaragiuntininegri2015b} Dalla Chiara, M. L., Giuntini, R., Leporini, R., Negri, E. and Sergioli, G. (2015). Quantum information, cognition, and music. {\it Frontiers in Psychology 6}, 1583. 

\bibitem{melucci2015} Melucci, M. (2015). {\it Introduction to Information Retrieval and Quantum Mechanics}. Berlin Heidelberg: Springer.

\bibitem{aertssozzoveloz2016} Aerts, D., Sozzo, S., and Veloz, T. (2016). A new fundamental evidence of non-classical structure in the combination of natural concepts. {\it Philosophical Transactions of the Royal Society A 374}, 20150095.

\bibitem{blutnerbeimgraben2016} Blutner, R. and beim Graben, P. (2016). Quantum cognition and bounded rationality. {\it Synthese 193}, 3239--3291.

\bibitem{broekaertetal2017} Broekaert, J., Basieva, I., Blasiak, P. and Pothos, E. M. (2017). Quantum-like dynamics applied to cognition: a consideration of available options. {\it Philosophical Transactions of the Royal Society A 375}, 20160387.

\bibitem{aertsetal2018a} Aerts, D., Aerts Argu\"elles, J., Beltran, L., Beltran, L., Distrito, I., Sassoli de Bianchi, M., Sozzo, S. and Veloz, T. (2018a). Towards a Quantum World Wide Web. {\it Theoretical Computer Science 752}, 116--131.

\bibitem{aertsbeltran2020} Aerts, D. and Beltran, L. (2020). Quantum structure in cognition: Human language as a Boson gas of entangled words. {\it Foundations of Science 25}, 755--802.

\bibitem{beltran2021} Beltran, L. (2021). Quantum Bose Einstein statistics for indistinguishable concepts in human language. {\it Foundations of Science}, doi 10.1007/s10699-021-09794-1.

\bibitem{aertsbeltran2022a} Aerts, D. and Beltran, L. (2022a). Are words the quantua of human language? Extending the domain of quantum cognition. {\it Entropy 24}, 6.

\bibitem{aertsbeltran2022b} Aerts, D. and Beltran, L. (2022b). A Planck radiation and quantization scheme for human cognition and language. {\it Frontiers in Psychology 13}, 8507255.

\bibitem{ijtp2023} Aerts, D., Aerts Argu\"{e}lles, J., Beltran, L., Geriente, S. and Sozzo, S. (2023). Entanglement as a method to reduce uncertainty. {\it International Journal of Theoretical Physics 62}, 145. 

\bibitem{philtransa2023} Aerts, D., Aerts Argu\"{e}lles J., Beltran, L., Geriente, S. and Sozzo, S. (2023). Development of a thermodynamics of human cognition and human culture. {\it Philosophical Transactions of the Royal Society A}, 10.1098/rsta.2022.0378.

\bibitem{aerts2009b} Aerts, D. (2009b). Quantum particles as conceptual entities: A possible explanatory framework for quantum theory. {\it Foundations of Science 14}, 361--411.

\bibitem{aertsetal2020} Aerts, D., Sassoli de Bianchi, M., Sozzo, S. and Veloz, T. (2020). On the conceptuality interpretation of quantum and relativity theories. {\it Foundations of Science 25}, 5--54.

\bibitem{moskovchenkofrenkel1990} Moskovchenko, N. I. and Frenkel, V. (Eds.). (1990). Ehrenfest-Ioffe. Nauchnaya Perepiska 1907--1933 gg. Leningrad: Nauka.

\bibitem{temperature2024} Aerts, D., Aerts Argu\"{e}lles, J., Beltran, L. and Sozzo, S. (2024). A thermodynamics for human cognition and language: Introducing Temperature. In preparation. 

\bibitem{eisbergresnick1985} Eisberg, R. and Resnick, R. (1985). {\it Quantum Physics of Atoms, Molecules, Solids, Nuclei, and Particles}. New York: John Wiley \& Sons.

\bibitem{pauli1940} Pauli, W. (1940). The Connection Between Spin and Statistics. {\it Physical Review 58}, 716--722.

\bibitem{cornellwieman2002} Cornell, E. A. and Wieman, C. E. (2002). Nobel Lecture: Bose--Einstein condensation in a dilute gas, the first 70 years and some recent experiments. {\it Reviews of Modern Physics 74}, 875.

\bibitem{ketterle2002} Ketterle, W. (2002). Nobel lecture: When atoms behave as waves: Bose--Einstein condensation and the atom laser. {\it Reviews of Modern Physics 74}, 1131.

\bibitem{annett2005} Annett, J. F. (2005). {\it Superconductivity, Superfluids, and Condensates}. Oxford: Oxford University Press. 

\bibitem{diekslubberdink2020} Dieks, D. and Lubberdink, A. (2020). Identical quantum particles as distinguishable objects. {\it Journal for General Philosophy of Science 53}, 259--274.

\bibitem{frenchredhead1988} French, S. and Redhead, M. (1988). Quantum physics and the identity of indiscernibles. {\it The British Journal for the Philosophy of Science 39}, 233--246.

\bibitem{saunders2003} Saunders, S. (2003). Physics and Leibniz's principles. In K. Brading and E. Castellani (Eds.), {\it Symmetries in Physics: Philosophical Reflections}, pp. 289--307. Cambridge, UK: Cambridge University Press.

\bibitem{mullerseevinck2009} Muller, F. A. and Seevinck, M. (2009). Discerning elementary particles. {\it Philosophy of Science 76} 179--200.

\bibitem{krause2010} Krause, D. (2010). Logical aspects of quantum (non-)individuality. {\it Foundations of Science 15}, 79--94.

\bibitem{wien1897} Wien, W. (1897). XXX. On the division of energy in the emission-spectrum of a black body. {\it The London, Edinburgh, and Dublin Philosophical Magazine and Journal of Science 43}, 214--220.

\bibitem{rubenskurlbaum1900} Rubens, H. and Kurlbaum, S. B. (1900). \"Uber die Emission langwelliger W\"armestrahlen durch den schwarzen K\"orper bei verschiedenen Temperaturen. {\it Sitzungsberichte der Preussischen Akademie der Wissenschaften, Physikalisch-mathematische Klasse}, 929--941.

\bibitem{boltzmann1884} Boltzmann, L. (1884). Ableitung des Stefan'schen Gesetzes, betreffend die Abh\"angigkeit der W\"armestrahlung lvon der Temperatur aus der electromagnetischen Lichttheorie. {\it Annalen der Physik 258}, 291-294. 

\bibitem{debroglie1924} de Broglie, L. (1924). Recherche sur la Th\'eorie des Quanta. Ph.D. Thesis, Facult\'e des sciences de Paris, Paris, France.

\bibitem{boltzmann1878} Boltzmann, L. (1877). Ũber die Beziehung zwischen dem zweiten Hauptsatze der mechanischen Wärmetheorie und der Wahrscheinlichkeitsrechnung respektive den Sätzen über das Wärmegleichgewicht. {\it Wiener Berichte 76}, 373--435. 

\bibitem{gibbs1902} Gibbs, J. W. (1902). {\it Elementary Principles in Statistical Mechanics}. Dover: New York.

\bibitem{planck1925} Planck, M. (1925). Zur Frage der Quantelung einatomiger Gase. {\it Sitzungsberichte der Preussischen Akadamie der Wissenschaften} 49--57.

\bibitem{schrodinger1925} Schrödinger, E. (1925). Bemerkungen über die statistische Entropiedefinition beim idealen Gas. {\it Sitzungsberichte der Preussischen Akademie der Wissenschaften 1925}, 434--441.

\bibitem{dirac1925} Dirac, P. A. M. (1925). Einstein--Bose statistical mechanics. Unpublished manuscript. In {\it Archives for History of Quantum Mechanics}, Microfilm 36, Sect. 9., p. 7.

\bibitem{schrodinger1989} Schrödinger, E. (1989). {\it Statistical Thermodynamics}. Mineola: Dover Publications.

\bibitem{dieks2011} Dieks, D. (2011). The Gibbs paradox revisited. In D. Dieks, W. Gonzalez, S. Hartmann. T. Uebel and M. Weber (Eds.), {\it Explanation, Prediction and Confirmation. The Philosophy of Science in a European Perspective}, pp. 367--377. Springer: Dordrecht.


\bibitem{schrodinger1926b} Schr\"odinger, E. (1926b). Quantisierung als Eigenwertproblem. {\it Annalen der Physik 384}, 361--376. 

\bibitem{aertssozzoveloz2015} Aerts, D., Sozzo, S. and Veloz, T. (2015). The quantum nature of identity in human thought: Bose--Einstein statistics for conceptual indistinguishability. {\it International Journal of Theoretical Physics 54}, 4430--4443.

\bibitem{milne1926} Milne, A. A. (1926). {\it Winnie-the-Pooh}. London: Methuen \& Co. Ltd.

\bibitem{zipf1935} Zipf, G. K. (1935). {\it The Psycho-Biology of Language}. Boston: Houghton Mifflin Co.

\bibitem{zipf1949} Zipf, G. K. (1949). {\it Human Behavior and the Principle of Least Effort}. Cambridge: Addison Wesley.

\bibitem{aertssassoli2014} Aerts, D. and Sassoli de Bianchi, M. (2014).The Extended Bloch Representation of Quantum Mechanics and the Hidden-Measurement Solution to the Measurement Problem. {\it Annals of Physics 351}, 975--1025. 

\bibitem{aertssassoli2015} Aerts, D. and Sassoli de Bianchi, M. (2015). The unreasonable success of quantum probability II: Quantum measurements as universal measurements. {\it Journal Mathematical Psychology 67}, 76--90.

\bibitem{italiancorpora2024} Aerts, D., Aerts Argu\"{e}lles, J., Beltran, L. and Sozzo, S. (2024). Identifying quantum mechanical statistics in Italian corpora. In preparation. 

\bibitem{aertsetalPhilTransA2024} Aerts, D., Aerts Argu\"{e}lles, J., Beltran, L., Sassoli de Bianchi, M. and Sozzo, S. (2024). The separability problem in quantum mechanics: Insights from research on axiomatics and human language. To be submitted to {\it Philosophical Transactions of the Royal Society A}. 

\bibitem{gemmeretal2009} Gemmer, J., Michel, M. and Mahler, G. (2009). {\it Quantum Thermodynamics: Emergence of Thermodynamic Behavior Within Composite Quantum Systems}. Lecture Notes in Physics 784. Berlin: Springer.

\bibitem{mahler2015} Mahler, G. (2015). {\it Quantum Thermodynamic Processes}. Singapore: Pan Stanford.

\bibitem{gogolineisert2016} Gogolin, C. and Eisert, J. (2016). Equilibration, thermalisation, and the emergence of statistical mechanics in closed quantum systems. {\it Reports on Progress in Physics 79}, 056001.


\end{thebibliography}
\end{document}